\documentclass{aa}
\usepackage{txfonts}
\usepackage[applemac]{inputenc}
\usepackage{natbib}
\usepackage{graphicx}
\usepackage{mathrsfs}

\bibpunct{(}{)}{;}{a}{}{,}
\usepackage{natbib}

\usepackage{subfig} 

\newcommand{\cch}{$\mathrm{C_{2}H}$}
\newcommand{\nnh}{$\mathrm{N_{2}H^{+}}$}
\newcommand{\co}{$\mathrm{C^{18}O}$}

\begin{document}

\title{C$_{2}$H in prestellar cores}

\author{M. Padovani\inst{1,2}, C.M. Walmsley\inst{2}, M. Tafalla\inst{3}, 
D. Galli\inst{2} \and H.S.P. M\"uller\inst{4}}

\institute{
Universit\`a di Firenze, Dipartimento di Astronomia e Scienza dello 
Spazio, Largo E. Fermi 2, I--50125 Firenze, Italy
\and 
INAF-Osservatorio Astrofisico di Arcetri, Largo E. Fermi 5, I--50125 
Firenze, Italy\\
\email{padovani@arcetri.astro.it, walmsley@arcetri.astro.it, 
galli@arcetri.astro.it}
\and 
Observatorio Astron\'omico Nacional, Alfonso XII 3, E--28014 Madrid, 
Spain\\
\email{m.tafalla@oan.es}
\and
I. Physikalisches Institut, Universit\"at zu K\"oln, Z\"ulpicher 
Stra{\ss}e 77, 50937 K\"oln, Germany\\
\email{hspm@ph1.uni-koeln.de}
}

\date{Received <date> / Accepted <date>}

\abstract {} 
{We study the
abundance of \cch\ in prestellar
cores both because of its role in the chemistry and because it
is a potential probe of the magnetic field.  We also consider
the non-LTE behaviour 
of the $N$=1$-$0 and $N$=2$-$1 transitions of \cch\
and improve current estimates of the spectroscopic
constants of \cch.}
{We used the IRAM 30m radiotelescope to map
the $N$=1$-$0 and $N$=2$-$1 transitions of \cch\ towards the prestellar cores
L1498 and CB246.  Towards CB246, we also mapped the 1.3 mm
dust emission, the $J$=1$-$0 transition of \nnh\ and the $J$=2$-$1 transition of 
\co. We used
a Monte Carlo radiative transfer program to analyse the
\cch\ observations of L1498. We derived the distribution of \cch\
column densities and compared with the H$_{2}$ column densities
inferred from dust emission.}
{We find that while non-LTE intensity ratios of different components 
of the $N$=1$-$0 and $N$=2$-$1 lines are  present, they are 
of minor importance and do not impede \cch\ column density 
determinations based upon
LTE analysis. Moreover, the comparison of our Monte-Carlo calculations 
with
observations suggest that the non-LTE deviations can be qualitatively
understood. For extinctions less than 20 visual magnitudes, we
derive toward these two cores (assuming LTE)
a relative abundance [\cch]/[H$_{2}$] of
$(1.0\pm0.3)\times10^{-8}$ in L1498  and $(0.9\pm0.3)\times10^{-8}$ in 
CB246 in reasonable agreement with our Monte-Carlo estimates. For
L1498, our observations in conjunction with the Monte Carlo
code imply a \cch\ depletion hole of radius $9\times10^{16}$ cm similar
to that found for other C-containing species.  We briefly discuss
the significance of the observed \cch\ abundance distribution. Finally, 
we used our observations
to provide improved estimates for the rest frequencies of all
six components of the \cch~(1$-$0) line and seven components of
\cch~(2$-$1).  Based on these results, we compute improved
spectroscopic constants for \cch. We also give a brief discussion
of the prospects for measuring magnetic field strengths using \cch.}
{}


\maketitle

\section{Introduction} 
The ethynyl radical \cch\ is a crucial intermediate in the interstellar 
chemistry 
leading to long chain carbon compounds. It has a $^{2}\Sigma$ ground 
state 
giving rise to non-negligible Zeeman splitting \citep{bl98}. 
These facts make it of interest to explore its abundance variations in 
nearby prestellar cores.
In particular, it seems worthwhile to test for its presence or absence 
in nearby cores where CO appears to be depleted \citep[see][]{bt07}. 
In most circumstances, CO is the main reservoir of C and thus depletion 
of CO seems 
likely to cause a reduction in the formation rate of C-containing 
species. 
However, recently \citet{hw08} have shown that in L1544 
and L183, CN appears to be present in the high density regions 
where CO is depleted.  This suggests that in some cases, CO depletion
may be ``modest'' in the sense that it causes the  high density
core nucleus to be invisible in CO isotopomers but the depletion is 
not sufficient to substantially change the abundances of minor C-
containing species.
In view of these facts, it seems reasonable to examine the behaviour
of other C-containing species in cores.

\cch\, is also of interest because hyperfine and spin interactions cause 
the rotational transitions to split into as many as 11 components which
can be observed simultaneously with modern autocorrelation spectrometers.
This allows for a precise examination of the deviations from LTE in 
individual rotational levels and eventually an evaluation of the 
relative importance of collisional and radiative processes.
One can observe for example in the $N$=1$-$0, 3 mm, transition six 
components 
with line strengths varying by over an order of magnitude and whose 
relative intensities 
can be compared extremely precisely. 
Departures from LTE have long plagued column density estimates in 
molecular 
lines which are often uncertain by a factor of order 2 as a
consequence. Reducing such uncertainties can only be achieved 
by understanding the causes of non-LTE behaviour in level populations
and this requires both theoretical (calculations of collisional
rates) and observational work. In this study, we attempt
to delineate the problem in the case of \cch\, from an observational 
point of view. 

We have chosen to study two cores with contrasting properties. L1498
is a well studied core in the Taurus complex at a distance of 140 parsec 
with a clear CO depletion hole.  Its density structure has been 
studied in detail 
by \citet{sn95} who conclude that their results are consistent with a 
``Bonnor-Ebert sphere'' of central  density $1-3\times10^4$ cm$^{-3}$ and 
by Tafalla et al.~(2004, 2006) who find a considerably higher central 
density 
of $9\times10^4$ cm$^{-3}$.  \citet{kc06} used SCUBA polarisation 
measurements 
to infer a surprisingly low value of the magnetic field of $10\pm 7$ $\mu
$G 
in the plane of the sky.
\citet{ah05} modelled the molecular distribution and concluded 
that their results were consistent with the contraction of a
``near equilibrium'' core. There is in general evidence for depletion of
C-bearing species such as $c$-C$_{3}$H$_{2}$ and C$_{2}$S in a central 
hole of 
radius $10^{17}$ cm \citep{tm06}. 
This is in contrast to NH$_{3}$ and N$_{2}$H$^{+}$ which show no signs of 
depletion 
in the central high density region of the core. 

CB246 (L1253) is a relatively isolated globule without an
associated IRAS source at a distance of 140$-$300 parsec 
(Dame et al.~1987, Launhardt \& Henning 1997) in the general direction of 
the
Cepheus flare. For the purpose of this study, we adopt a distance of 200 
pc. 
CB246 is apparently a double core seen both in
NH$_{3}$ and C$_{2}$S 
(Lemme et al.~1996, Codella \& Scappini~1998) on a size scale of roughly 
0.1 parsec.  
The mass (dependent on the distance) is in the range 0.2$-$1 M$_{\sun}$ 
\citep{cs98} from the NH$_{3}$ maps, though there is
considerable uncertainty in this.  From a chemical point of view, it is
interesting that there is rough general agreement between the spatial
distributions seen in NH$_{3}$ and C$_{2}$S. One aim of the present 
observations has been to check if \cch\, shows signs of depletion towards 
the peak emission seen in NH$_{3}$.

In this article, we present IRAM 30m maps of the emission in the $N$=1$-
$0
transition of \cch\, towards CB246 and L1498. We supplement this with
measurements at selected positions of the $N$=2$-$1 transition of \cch\, 
as 
well as maps of the 1.3 mm dust emission, the $J$=1$-$0 transition of
\nnh\ and the $J$=2$-$1 transition of \co\ 
towards CB246. 
In Section 2, we describe our observational and data reduction procedures 
and in Section
3, we summarise the observational results from the line measurements. In 
Section 4,  
we attempt to use our astronomical observations to estimate rest 
frequencies 
for the individual components of \cch(1$-$0) and (2$-$1) and an updated 
set of hyperfine parameters.
In Section 5, we give our conclusions concerning the deviations from
LTE populations for \cch\, as well as a very tentative interpretation. 
In Section 6, we give our column density and abundance estimates in the 
two objects and
discuss the evidence for depletion of \cch\,.  
In Section 7, we discuss our results and in Section 8, we summarise our 
conclusions.

\section{Observations}

\subsection{C$_{2}$H in L1498 and CB246}
The observations were carried out with the IRAM 30m telescope. 
The \cch(1$-$0) multiplet (at 87 GHz: see Table \ref{tab:c2h10p}) 
was observed at different epochs between August 2007 (only component 2, 
3, 4 and 5) 
and July 2008 (all the components), with 1$-$2 mm precipitable water 
vapor (pwv). 
Observations of the \cch(2$-$1) multiplet 
(at 174 GHz: see Table \ref{tab:c2h21p}) were performed in July 2008 with 
3$-$4 mm pwv.
The HPBW at the \cch(1$-$0) and \cch(2$-$1) frequencies are 28\arcsec\ 
and 14\arcsec, respectively.
In August 2007, the VESPA autocorrelator was used to obtain 10 kHz 
channel spacing (corresponding to about 0.034 km s$^{-1}$)
with 20 and 40 MHz bandwidths, while in July 2008, the autocorrelator was 
used 
to obtain 20 kHz channel spacing with 20 and 40 MHz bandwidths for 
\cch(1$-$0) 
and 40 kHz channel spacing with 40 and 80 MHz bandwidths for \cch(2$-$1).
In this way we covered the six hyperfine structure (HFS) components of 
\cch(1$-$0) 
and seven HFS components of \cch(2$-$1).
 
The two cores, L1498 and CB246, were mapped in \cch(1$-$0) in 2008
in raster mode with a spacing of 20\arcsec\ (channel spacing 20 kHz)
for L1498 and of 15\arcsec\ for CB246 (thus close to Nyquist sampling for
CB246). In 2008, we also observed the \cch(2$-$1) line at the
positions given in Table \ref{tab:c2h21}. Finally in 2007, we observed
the \cch(1$-$0) line towards the (0,0) offset with 10 kHz resolution in
both sources.
 
The observed strategy was identical for all measurements: 
frequency-switching mode 
with a 7.5 MHz throw and a phase time of 0.5 s, with a calibration every 
10 to 15 minutes.
The data were reduced using CLASS, the line data analysis program of the 
GILDAS 
software\footnote{\tt http://www.iram.fr/IRAMFR/GILDAS.}.
Instrumental bandpass and atmospheric contributions were subtracted with 
polynomial baselines, 
before and after the folding of the two-phase spectra.
The final rms, in the main-beam temperature ($T_{\mathrm{mb}}$), 
in each channel of width $\delta v=0.067$ km s$^{-1}$ is $
\sigma_{\mathrm{T}}\sim50$ mK for 
both L1498 and CB246 and for both the $N$=1$-$0 and the $N$=2$-$1 
transitions, while the system temperatures are $T_{\mathrm{sys}}
\sim130$ K 
for the $N$=1$-$0 transition and $T_{\mathrm{sys}}\sim1000$ K for the
$N$=2$-$1 
transition.
In what follows, all temperatures are on the main-beam scale, 
$T_{\mathrm{mb}}=F_{\mathrm{eff}}T_{\mathrm{A}}^{*}/B_{\mathrm{eff}}$, 
where $T_{\rm A}^{*}$ is the antenna temperature corrected for 
atmospheric absorption, 
and the forward and beam efficiencies are respectively 
$F_{\mathrm{eff}}=0.95$ 
and $B_{\mathrm{eff}}=0.77$ concerning the $N$=1$-$0 transition and 
$F_{\mathrm{eff}}=0.93$ 
and $B_{\mathrm{eff}}=0.65$ for the $N$=2$-$1 transition.

\subsection{N$_{2}$H$^{+}$(1$-$0) and C$^{18}$O(2$-$1) in CB246}
\label{nnh_coCB}
Observations of the \nnh(1$-$0) multiplet (at 93 GHz) and ${\rm C}^{18}
{\rm O}$(2$-$1) (at 219 GHz) in CB246 were carried out simultaneously 
in July 2008, 
 with 3$-$4 mm pwv. The HPBW at the \nnh(1$-$0) and C$^{18}$O(2$-$1) 
frequencies are 26\arcsec\ and 11\arcsec\ respectively.  
The VESPA autocorrelator was used to obtain 10 kHz channel spacing  
with 40 MHz bandwidth for N$_{2}$H$^{+}$(1$-$0) and 20 kHz channel  
spacing with 40 MHz bandwidth for \co(2$-$1). We observed using  
the frequency-switching mode with a 7.5 MHz throw for N$_{2}$H$^{+} 
$(1$-$0) and a 15 MHz throw for \co(2$-$1) and a phase time of 0.5  
s with a calibration every 10 to 15 minutes. We observed a region of
2.5 arc minutes squared in extent with a spacing of 15\arcsec. Thus, the
N$_{2}$H$^{+}$ map is essentially Nyquist sampled whereas
\co\ is under-sampled (though the maps shown subsequently are
smoothed to a resolution of 26\arcsec). 

For N$_{2}$H$^{+}$, the final rms, in  main-beam  
temperature units ($T$$_{\rm mb}$), in  channels of width $\delta v=0.031$  
km s$^{-1}$ was $\sigma_{\rm T}\sim110$ mK, with a $T_{\rm sys}\sim160$  
K.  For \co, the final rms (channels  
of width $\delta v=0.027$ km s$^{-1}$) was $\sigma_{\rm T}\sim200$ mK,  
with a $T_{\rm sys}\sim400$ K. The forward and the beam efficiencies  
are respectively $F_{\rm eff}=0.95$ and $B_{\rm eff}=0.77$ for \nnh\, 
and $F_{\rm eff}=0.91$ and $B_{\rm eff}=0.55$ for \co.

\subsection{Bolometer map of CB246}
\label{boloCB}
CB246 was observed in the 1.3 mm continuum with the IRAM 30m telescope  
in May, November and December 2007. We used the MAMBO2 117-channel  
bolometer array in the on-the-fly mode with a scanning speed of 6\arcsec\  
s$^{-1}$, a wobbler period of 0.5 s and a wobbler throw of 70\arcsec,  
reaching an rms of 3 mJy/beam.
The bolometer central frequency is 250 GHz and the half power  
bandwidth (HPBW) at 1.3 mm is approximately 11\arcsec.

\section{Observational results}

\subsection{C$_{2}$H(1$-$0)}
\label{obsrescch10}
\begin{figure}[!h] 
\resizebox{\hsize}{!}{\includegraphics[angle=-90]{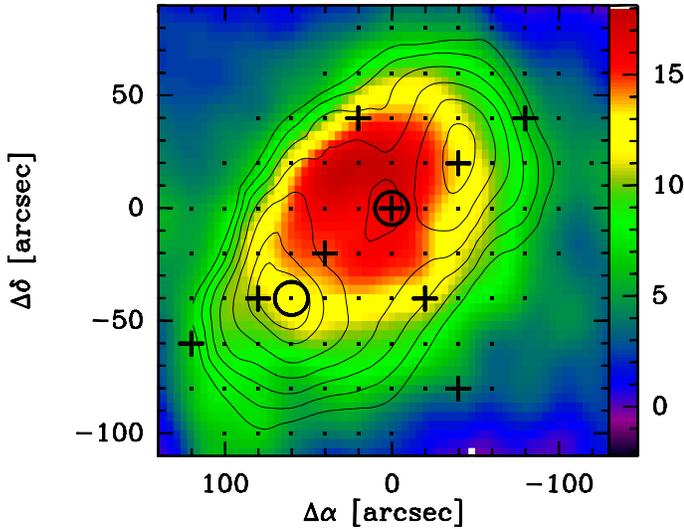}} 
\caption{Dust emission at 1.3 mm (after smoothing to 28\arcsec, {\em 
coloured map}) towards L1498, from Tafalla et al.~(2002), superposed to 
the emission of \cch(1$-$0) ($J',F'\rightarrow J,F = 
3/2,2\rightarrow1/2,1$), {\em black contours}. Contours represent 40, 50, 
60, 70, 80, 90 and 95 per cent of the peak value which is 0.67 K km s
$^{-1}$. The scale bar to the right of each panel gives the continuum 
flux in mJy/beam. The {\em crosses} indicate the positions used for the 
cuts in Fig. \ref{cut_paper} while the {\em circles} show positions 
observed in \cch(2$-$1). The (0,0) position corresponds to $
\alpha(2000)=04^{\rm h}10^{\rm m}51.5^{\rm s}$, $
\delta(2000)=25°09'58''$.}
\label{cont_C2H(2)L_crosses} 
\end{figure}

In Fig. \ref{cont_C2H(2)L_crosses}, we show our map of the integrated 
intensity 
in the \cch(1$-$0) ($J',F'\rightarrow J,F=3/2,2\rightarrow1/2,1$) 
transition 
superposed on the map of the 1.3 mm dust continuum emission 
smoothed at 28\arcsec\ together with the positions observed in the 
$N$=2$-$1 
transition.
One sees that towards L1498, while the dust emission has a single peak 
at the centre of the map, the \cch\,  emission has a broad ``plateau'' 
with an extent rather similar to that of the dust, showing two 
peaks to the SE and NW of the dust emission peak. 
Rather similar distributions have been seen in several other C-bearing 
species 
(e.g. CS, C$_{2}$S, CO and $c$-C$_{3}$H$_{2}$, see Tafalla et al.~2002 
and 
Tafalla et al.~2006) and has been attributed to depletion onto grain 
surfaces 
in the dense gas associated with the dust continuum peak. 
Our data suggest that \cch\, behaves in a similar fashion towards L1498.

\begin{figure}[!h] 
\resizebox{\hsize}{!}{\includegraphics[angle=-90]{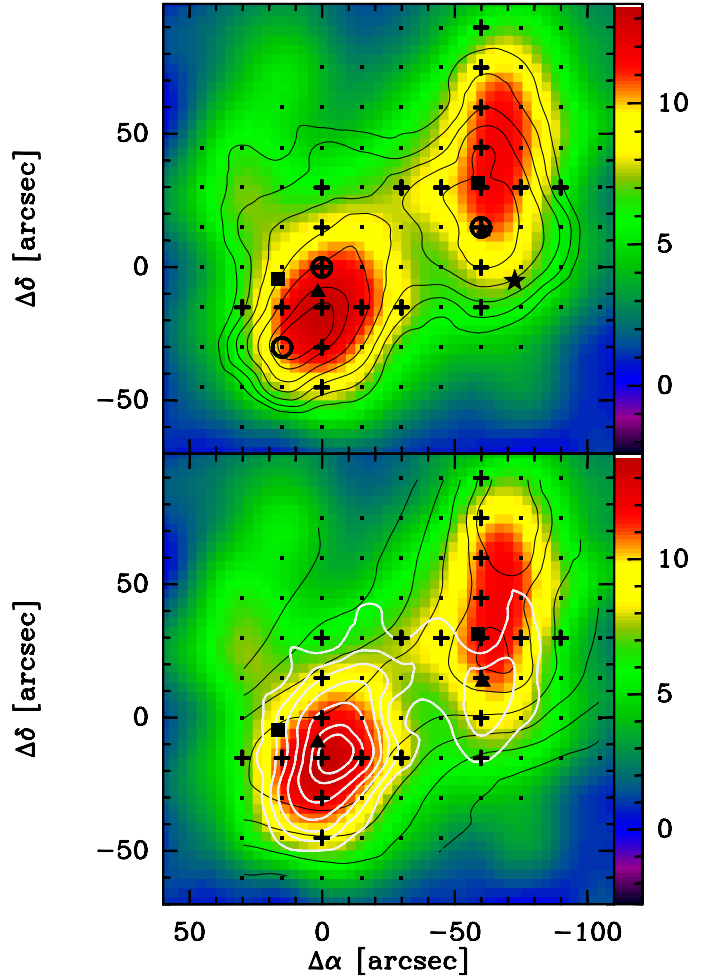}} 
\caption{Upper panel: dust emission at 1.3 mm (after smoothing to 
28\arcsec, {\em coloured map}) towards CB246 (see Sect. \ref{boloCB}), 
superposed to the emission of \cch(1$-$0) ($J',F'\rightarrow J,F = 
3/2,2\rightarrow1/2,1$); {\em circles} show positions observed in 
\cch(2$-$1); {\em asterisk} shows the position of the 2MASS object (see Sect. \ref{MassCB246}). Lower panel: dust emission at 1.3 mm (after smoothing to 
26\arcsec, {\em coloured map}) towards CB246 superposed to the emission 
of \co(2$-$1), {\em black contours}, and \nnh(1$-$0) 
($F_{1}',F'\rightarrow F_{1},F = 1,2\rightarrow0,1$), {\em white 
contours}; Both panel: contours represent 40, 50, 60, 70, 80, 90 and 95 
per cent of the peak values which are 0.67, 0.70 and 0.39 K km s$^{-1}$ 
for \cch(1$-$0), \co(2$-$1) and \nnh(1$-$0), respectively; the crosses 
indicate the positions used for the cuts in Fig. \ref{cut_paper} and 
\ref{cut_paper_appendix}; {\em solid triangles} represent the peaks of 
NH$_{3}$(1,1), from Lemme et al.~(1996), and {\em solid squares} the 
peaks 
of C$_{2}$S(2$_{1}$--1$_{0}$), from Codella \& Scappini (1998). The scale 
bar to the right of each panel gives the continuum flux in mJy/beam. The 
(0,0) position corresponds to $\alpha(2000)=23^{\rm h}56^{\rm m}43.6^{\rm 
s}$, $\delta(2000)=58°34'09''$.}
\label{CB_twomaps_paper} 
\end{figure}

The situation is rather different towards CB246 (see Fig. 
\ref{CB_twomaps_paper}, 
upper panel) where we see that the dust and \cch\, emission have rather 
similar distributions. 
Both are double peaked, although the \cch\, NW emission peak 
is offset 
about 30\arcsec\ to the south of its counterpart in dust emission whereas 
towards the SE peak the difference is marginal. 

\begin{figure}[!h] 
\resizebox{\hsize}{!}{\includegraphics[angle=-90]{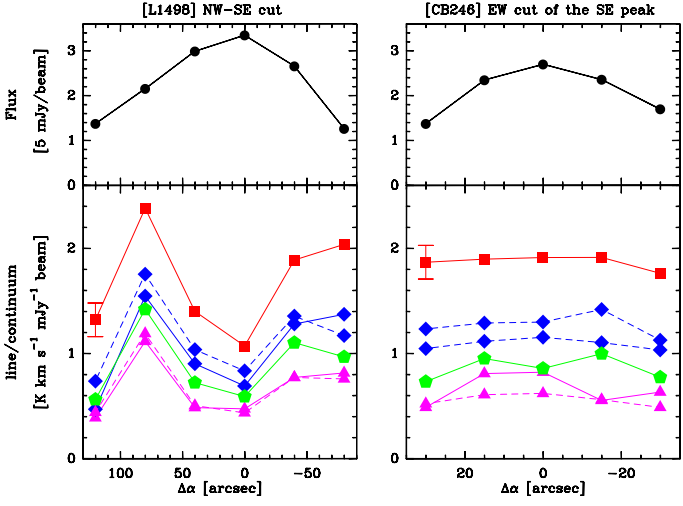}} 
\caption{Upper panels: Continuum emission flux measured along the NW-SE 
cut in L1498, 
see Fig. \ref{cont_C2H(2)L_crosses}, and along the EW cut of the SE peak 
in CB246, 
see upper panel of Fig. \ref{CB_twomaps_paper}. 
Lower panels: ratio between the integrated intensity of the different 
hyperfine components 
of \cch(1$-$0) with the continuum emission flux in the same positions of 
the upper panels. 
The typical error on the ratio is 0.16 K km s$^{-1}$ mJy$^{-1}$ beam 
(component 1, 
{\em solid magenta line with triangles}; component 2, {\em solid red line 
with squares}; 
component 3, {\em solid blue line with diamonds}; component 4, {\em 
dashed blue line with diamonds}; 
component 5, {\em solid green line with pentagons}; component 6, {\em 
dashed magenta line 
with triangles}; component labels are given in Table \ref{tab:c2h10p}).} 
\label{cut_paper} 
\end{figure}

It is interesting to compare the distributions of different components.
We do this in Fig. \ref{cut_paper} where we compare cuts in all the  
$1-0$ component lines 
with cuts in the continuum intensity. One sees that ratios of 
different pairs of components do not vary greatly 
along these cuts, although the ratio of the strong 87316 MHz component 
(no. 2 in 
Table \ref{tab:c2h10p}) to the weak 87284 MHz line (no. 1) at the L1498 
dust peak 
is somewhat smaller (about 45\%) than elsewhere in the cut. However, 
this ratio is always of order 3 
as compared to the expected value of 10 for optically thin emission
(suggesting a moderately optically thick 87316 MHz line).
There are thus indications of saturation of the stronger components of 
the \cch(1$-$0) line 
and we conclude that optical depth effects are present but lines 
have moderate opacities, as shown in Table \ref{tab:c2h10} for selected positions 
in the two observed sources. 
It is noticeable also that while \cch\, has a minimum towards the dust 
peak in L1498 
(and we conclude this is a real column density minimum), there is a 
little indication of 
variation in the ratio of \cch\, to dust continuum intensity crossing the 
SE peak in CB246. 

\begin{table}[!h]
\caption{Line parameters$^{a}$ observed in \cch(1$-$0)} 
\label{tab:c2h10} 
\centering 
\begin{tabular}{c c c c} 
\hline\hline 
$\alpha,\delta$ offsets & $V_\mathrm{LSR}$ & $\Delta V$ & $\tau_{\rm total}^{\;b}$ \\ 
$[$\arcsec,\arcsec] & [km s$^{-1}$] & [km s$^{-1}$] &\\  
\hline 
\multicolumn{4}{c}{L1498}\\
$0,0$ & 7.80(0.01) & 0.17(0.01) & 29.4(0.3)\\
$60,-40$ & 7.74(0.03) & 0.21(0.01) & 28.3(0.4)\\
\hline
\multicolumn{4}{c}{CB246}\\
$0,0$ & $-$0.83(0.01) & 0.24(0.01) & 13.5(0.3)\\
$15,-30$& $-$0.78(0.01) & 0.26(0.01) & 20.9(1.1)\\
$-60,15$ & $-$0.87(0.01) & 0.27(0.01) & 14.4(0.7)\\
\hline 
\end{tabular}\\[2pt]
$^a$ Numbers in parentheses represent the errors from the fit.\\
$^b$ Sum of the peak optical depth of the six hyperfine components.
\end{table}


\begin{figure}[!h] 
\resizebox{\hsize}{!}{\includegraphics[angle=0]{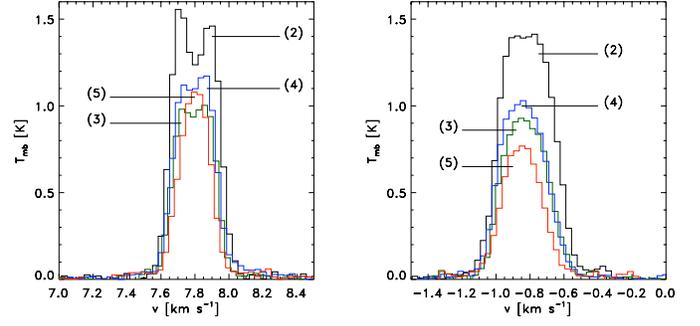}} 
\caption{Component 2 ({\em black}), 3 ({\em green}), 4 ({\em blue}) and 5 ({\em red}) of 
\cch(1$-$0) observed at high resolution in 2007 at the offset (0,0) in L1498 (left panel) 
and CB246 (right panel). In order to compare line shapes and intensities, components have been shifted in frequency. Component labels and rest frequencies are given in column 1 and 3 
of Table \ref{tab:c2h10p}).} 
\label{c2h_2007obs} 
\end{figure}

Another indication of line transfer effects can be obtained from 
comparing line profiles of the different \cch(1$-$0) components as shown 
in Fig. \ref{c2h_2007obs} where we show the high (10 kHz) spectral 
resolution data from 2007. 
It is noticeable that towards the dust peak, the weakest 87407 MHz 
component 
(no. 5 in Table \ref{tab:c2h10p}) has a maximum at a velocity where the 
strong 
87316 MHz line (no. 2) shows 
a dip between two peaks. This is the signature expected for ``self-
absorption'' 
by a foreground layer of density lower than that responsible for the bulk 
of the emission. 
The effect however is not sufficiently strong to change 
the above conclusions concerning the importance of optical depth effects. 
We note moreover that towards CB246, no effects of this type are seen.

\subsection{C$_{2}$H(2$-$1)}
\begin{figure}[!h] 
\centering 
\subfloat
{\includegraphics[width=.48\columnwidth,angle=0]{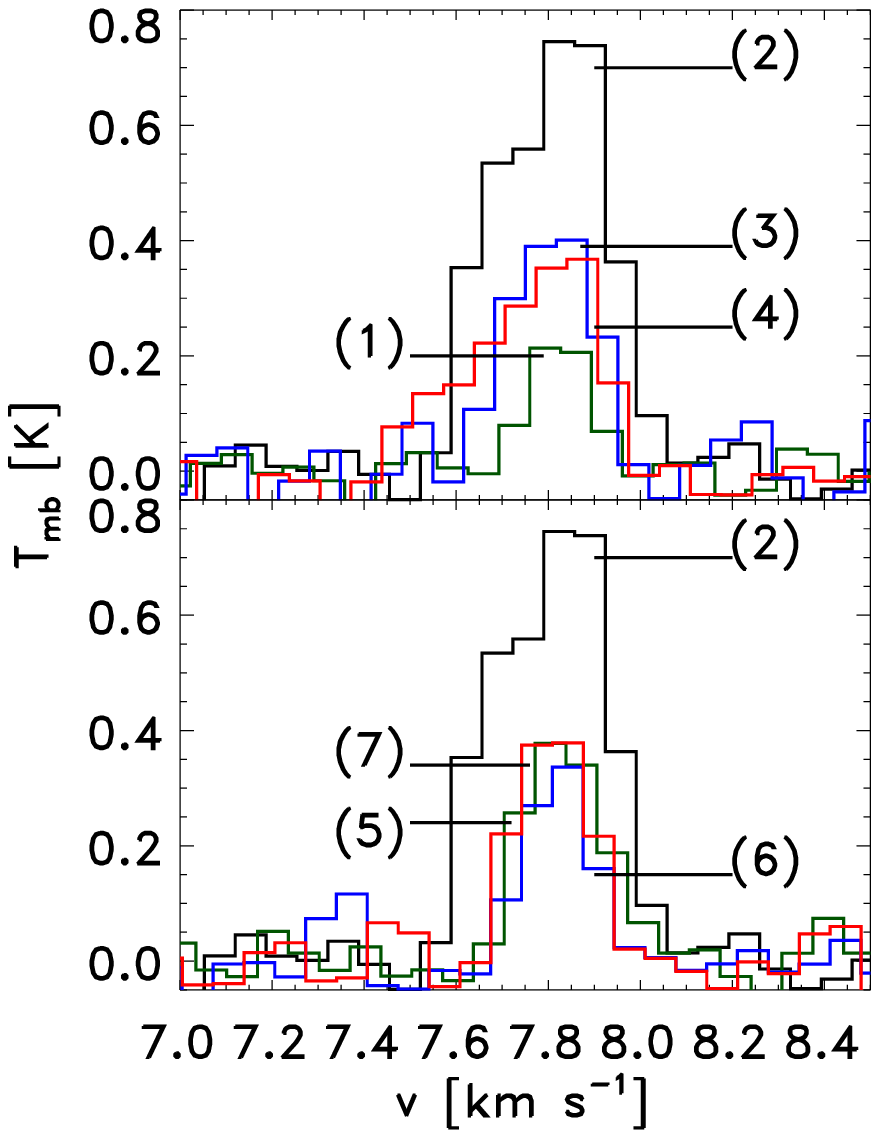}} \quad 
\subfloat 
{\includegraphics[width=.48\columnwidth,angle=0]{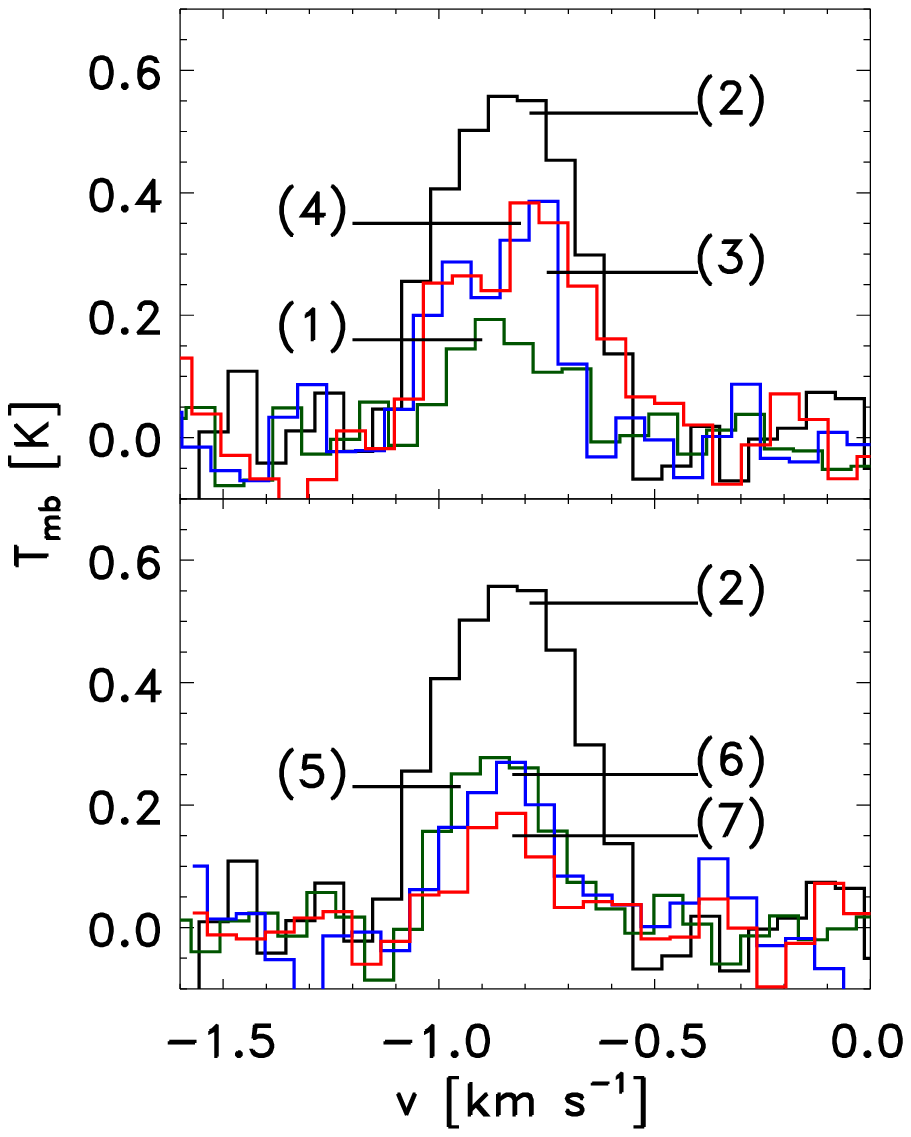}}
\\ 
\caption{Hyperfine components of \cch(2$-$1) observed at the offset (0,0) in L1498 
(left panel) and CB246 (right panel). In order to compare line shapes and intensities, components have been shifted in frequency. Component labels and rest frequencies are given 
in column 1 and 3 of Table \ref{tab:c2h21p}).} 
\label{c2h21_2008obs} 
\end{figure}

The \cch(2$-$1) line has had little (if any) attention and it was thus interesting that we succeeded in detecting 
7 of the 11 components of the line in both sources. 
Line parameters are given in Table \ref{tab:c2h21} and sample profiles are shown in 
Fig. \ref{c2h21_2008obs}. It is interesting (see discussion below) that within the errors, 
line intensities are consistent with LTE but that, in particular towards L1498, 
the total optical depth derived from an LTE fit (see Table \ref{tab:c2h21}) is large and corresponds 
to an optical depth of 3.1 in the strongest 174663 MHz component (no. 2 of Table \ref{tab:c2h21p}).
We note however that slight errors in rest frequencies could influence this interpretation.


\begin{table}[!h]
\caption{Line parameters$^{a}$ observed in \cch(2$-$1)} 
\label{tab:c2h21} 
\centering 
\begin{tabular}{c c c c} 
\hline\hline 
$\alpha,\delta$ offsets & $V_\mathrm{LSR}$ & $\Delta V$ & $\tau_{\rm total}^{\;b}$ \\ 
$[$\arcsec,\arcsec] & [km s$^{-1}$] & [km s$^{-1}$] &\\  
\hline 
\multicolumn{4}{c}{L1498}\\
$0,0$ & 7.84(0.01) & 0.23(0.01) & 8.6(1.8)\\
$60,-40$ & 7.76(0.01) & 0.25(0.01) & 9.7(1.9)\\
\hline
\multicolumn{4}{c}{CB246}\\
$0,0$ & $-$0.84(0.01) & 0.35(0.02) & 1.5(1.2)\\
$15,-30$& $-$0.73(0.01) & 0.30(0.01) & 5.4(1.0)\\
$-60,15$ & $-$0.88(0.01) & 0.30(0.01) & 5.5(1.2)\\
\hline 
\end{tabular}\\[2pt]
$^a$ Numbers in parentheses represent the errors from the fit.\\
$^b$ Sum of the peak optical depth of the seven hyperfine components.
\end{table}


\subsection{C$^{18}$O and N$_{2}$H$^{+}$}
 In the lower panel of Fig. \ref{CB_twomaps_paper}, we show a superposition 
of our 1.3 mm continuum map (smoothed to an angular resolution of 26\arcsec) 
with the \co(2$-$1) and the isolated component of \nnh(1$-$0) 
($F'_{1},F'\rightarrow F_{1},F=0,1\rightarrow1,2$) integrated intensity maps, 
together with the emission peaks in NH$_{3}$(1,1), using the data obtained by 
Lemme et al.~(1996) and in C$_{2}$S($2_{1}-1_{0}$) 
by Codella \& Scappini~(1998).  There is a general similarity between the distributions of \co\ 
seen here and that of \cch\, (Fig. \ref{CB_twomaps_paper}, upper panel). 
In general, also the \co\ distribution resembles that seen in the continuum 
but there are clear shifts (for example in the SE) between the peaks seen in the dust emission 
and in \co.  In the NW, also the ``bar''-like structure seen in the continuum to be extended 
N-S is also present in \co\ but seems even more elongated (dimensions of  100\arcsec\ 
in the continuum as compared to 120\arcsec\ in \co).
  
To the SE there is reasonable agreement between the continuum and \nnh\ 
intensity distributions and the ammonia peak appears to be consistent with these. 
To the NW on the other hand, the continuum is extended in a ``bar''-like feature 
and \nnh\ (but also \cch) peaks at the south end of this 40\arcsec\ to the south of 
the continuum peak. These differences suggest chemical differentiation in CB246 albeit 
on a scale smaller than in L1498. Besides, the \nnh\ peaks resemble the continuum structure 
in the SE but differ considerably in the NW suggesting that in this case 
(in contrast to L1498 and L1544, see Tafalla et al.~2006), 
the \nnh\ abundance may be varying with position.
The shift between the \nnh\ NW peak and the continuum peak is  
unusual, but not unprecedented (see Pagani et al.~2007).
However, a more detailed study of the radiative transfer and excitation 
is needed to confirm this.

\begin{figure}[!h] 
\resizebox{\hsize}{!}{\includegraphics[angle=-90]{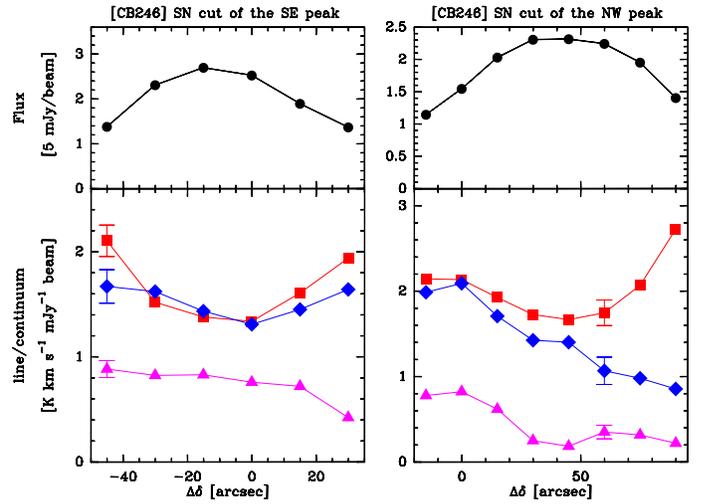}} 
\caption{Upper panels: Continuum emission flux measured along the SN cut of the SE 
and the NW peaks in CB246, see upper panel of Fig. \ref{CB_twomaps_paper}. 
Lower panels: ratio between the integrated intensity of the component 2 of 
\cch(1$-$0), {\em red squares}, \co(2$-$1), {\em blue diamonds}, and the isolated component 
of \nnh(1$-$0), {\em magenta triangles}, with the continuum emission flux 
in the same positions of the upper panels. The typical errors on the ratio are 0.16, 0.15 
and 0.08 K km s$^{-1}$ mJy$^{-1}$ beam for \cch, \co\ and \nnh\ respectively.} 
\label{cut_paper_appendix} 
\end{figure}

In Fig. \ref{cut_paper_appendix} we compare cuts in the strongest component of \cch(1$-$0) 
(component 2, see Table \ref{tab:c2h10p}), \co(2$-$1) and the isolated component 
of \nnh(1$-$0) with cuts in the continuum intensity. \nnh\ and \co\ show a rather constant 
line-to-continuum ratio in the SE peak cut, while in the NW peak cut, these molecules do not look 
like the continuum structure, especially the \co\ emission which appears to be shifted down 
compared to the continuum emission.


\subsection{Mass of CB246}
\label{MassCB246}
In the continuum, CB246 shows a double-peaked profile, with a rounded SE clump with a radius (measured at half-power contour) of 33\arcsec\ (corresponding to 0.03 pc at the distance of 200 pc) and a more elliptical NW clump with a major and a minor axes of 100\arcsec\ (or 0.1 pc) and 44\arcsec (or 0.04 pc), respectively. Dust emission is generally optically thin at millimetre wavelengths and hence is a direct tracer of the mass content of molecular cloud cores. In the hypothesis of an isothermal dust source, the total (dust plus gas) mass is related to the millimetre flux density, $S_{1.3}$, integrated over the solid angle $\Omega_{\rm beam}$ according to the following equation

\begin{equation}
M=\frac{S_{1.3}\,d^{2}}{\kappa_{1.3}B_{1.3}(T_{\rm d})}
\end{equation}
where $B_{1.3}(T_{\rm d})$ is the Planck function calculated at the dust 
temperature $T_{\rm d}=10$ K and $\kappa_{1.3}$ is the dust opacity per 
unit mass which is assumed constant and equal to 0.005  cm$^{2}$ g
$^{-1}$ for prestellar cores of intermediate densities ($n\lesssim10^{5}$ 
cm$^{-3}$), assuming a gas-to-dust mass ratio of 100 (Henning et 
al.~1995, Preibisch et al.~1993). We found a FWHM mass of 1.2 M$_{\odot}$ 
for both clumps and a total mass of 3.3 M$_{\odot}$ which has been 
evaluated taking into account the continuum flux above two times the rms.
The average column and number density of molecular hydrogen have
been evaluated within the same area used to compute the mass and
are to $N(\mathrm{H}_{2})=1.9\times10^{22}$ cm$^{-2}$ 
and $n(\mathrm{H}_{2})=7.6\times10^{4}$ cm$^{-3}$, respectively.

To the south of the NW emission peak, there is a bright infrared source (2MASS J23563433+5834043, see Fig.~\ref{CB_twomaps_paper}) which appears to be a heavily reddened background star with visual extinction of $\sim$ 20 magnitudes, consistent with our column density estimates.

\subsection{Gas kinematics}
\begin{figure}[!h]
\begin{center} 
\resizebox{\hsize}{!}{\includegraphics[angle=-90]{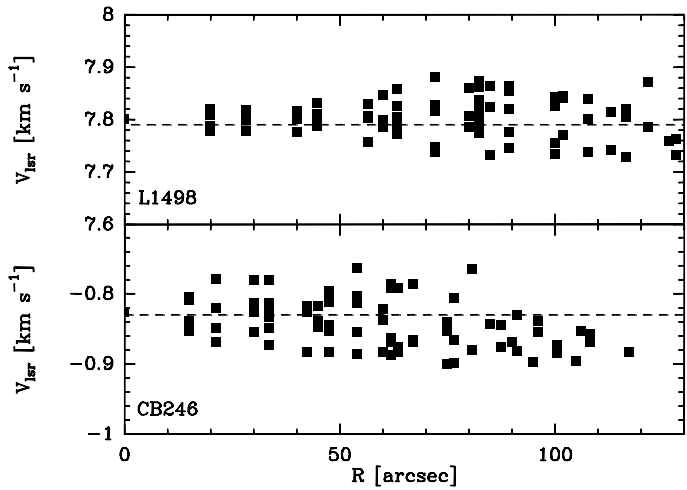}} 
\caption{Radial profile of line center velocity for L1498 (upper panel) 
and CB246 (lower panel) derived from the hyperfine-structure fits of 
\cch(1$-$0). The uncertainty of the velocity estimates is of the order of 
0.003 km s$^{-1}$ and the dashed line represents the mean value.}
\label{LCBvlsr}
\end{center} 
\end{figure}

From the hyperfine fit of \cch(1$-$0), using the rest frequencies discussed below (see Sect.~\ref{restfreq}), we derived the line center 	
velocities, $V_{\rm lsr}$, and in Fig. \ref{LCBvlsr} we present the 
radial profiles which are almost flat on the average, especially for 
L1498. Radii
are computed with respect to the offset (0,0) for the two sources (see 
Fig. \ref{cont_C2H(2)L_crosses} and \ref{CB_twomaps_paper}).

\begin{figure}[!h]
\begin{center} 
\resizebox{\hsize}{!}{\includegraphics[angle=-90]{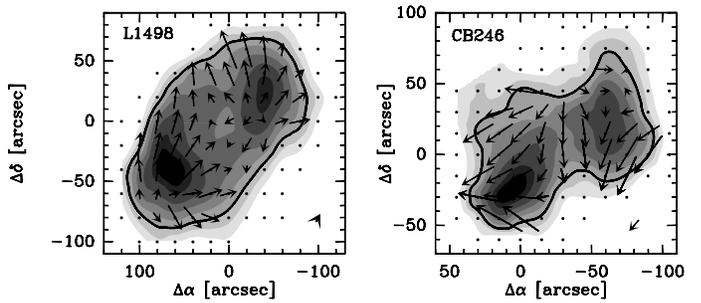}} 
\caption{\cch(1$-$0) velocity gradient toward L1498 (upper panel) and 
CB246 (lower panel): the integrated intensity map of component 2 of \cch\ 
is shown in gray scale (black contour represents the 50\% of the peak 
value) while arrows represent the local velocity gradients in the 
adjacent nine points pointing in the direction of increasing velocity and 
with the length proportional to the magnitude. The arrow in the bottom-
right corner represents the total velocity gradient.} 
\label{LCBgrad}
\end{center} 
\end{figure}

Following Goodman et al.~(1993), we checked for the presence of a 
velocity gradient, $\mathscr{G}$, across these two cores using \cch(1$-
$0) data. A further check has been done on CB246 using \nnh(1$-$0) and 
\co(2$-$1) data, finding a good agreement within the errors with the 
corresponding values from \cch. In Fig. \ref{LCBgrad} we show the local 
velocity gradients in the two cores and in Table \ref{tab:grad} we show 
our results as regards the velocity gradient, $\mathscr{G}$, its 
direction, $\Theta_{\vec{\mathscr{G}}}$ and the mean LSR velocity, $
\langle V_{\rm lsr}\rangle$. All these quantities are in good agreement 
with the estimates of Goodman et al.~(1993) and Tafalla et al.~(2004) who 
observed L1498 using NH$_{3}$ and \nnh\ as a tracer. Besides, we 
evaluated the angular velocity, $\omega$, assuming that the angular 
velocity vector points in the direction given by $\vec{\hat{\omega}}$, 
that is $\vec{\omega}=(\mathscr{G}/\sin i)\,\vec{\hat{\omega}}$, where $i
$ is the inclination of $\vec{\omega}$ to the line of sight and the 
position angle of $\vec{\hat{\omega}}$ is given by $
\Theta_{\vec{\hat{\omega}}}=\Theta_{\vec{\mathscr{G}}}+\pi/2$. 
Statistically, for a random distribution of orientation, $\langle\sin i
\rangle^{-1}=4/\pi$ (Chandrasekhar \& M\"unch 1950, Tassoul 1978).
To quantify the dynamic role of the rotation in a cloud, we calculated 
the ratio between the rotational and the gravitational energy, denoted 
with $\beta$. For a uniform density sphere, $\beta=(\omega^{2}R^{3})/
(3GM)$, where $R$ is the rotation radius, $G$ the gravitational constant 
and $M$ the mass, and the specific angular momentum is defined as $L/M=(2/5)R^{2}\omega$. 

The values of $\beta$ we found suggest that the clouds have little 
rotational energy as generally found in molecular cloud cores (e.g. 
Goodman et al.~1993). The specific angular momentum we evaluated for 
these two cores is of order $10^{-3}$ km s$^{-1}$ pc. In the hypothesis 
that a core has been formed from a parental clump with equal mass and at 
the same galactocentric distance of the core (about 7 kpc for L1498 and 
CB246), we computed a rotation frequency of order 10$^{-15}$ s$^{-1}$ 
assuming that the angular momentum of the clump is due to Galactic 
differential rotation (Clemens, 1985).
This value corresponds to a specific angular momentum of $3\times10^{-2}$ 
km s$^{-1}$ pc, that is 30 times greater than the $L/M$ of the cores.
This shows that the specific angular momentum of the cores is low 
compared to larger (galactic) scales. Thus angular momentum is lost in 
forming cores as expected by Ohashi (1999) for cores with radius greater 
than 0.03 pc. In particular, the cores lies on the relation founded by Ohashi (1999) for $L/M$ as a function of the radius.

\begin{table}[!h]
\caption{Results of gradient fitting.}
\begin{center}
\begin{tabular}{cccc}
\hline\hline
quantity & unit & L1498 & CB246\\
\hline
$\mathscr{G}$ & [km s$^{-1}$ pc$^{-1}$] & $0.56\pm0.05$ & $0.65\pm0.09$\\
$\Theta_{\vec{\mathscr{G}}}$ & [deg E of N] & $-37\pm1$ & $143\pm4$\\
$\langle V_{\rm lsr}\rangle$ & [km s$^{-1}$] & $7.79\pm0.04$ & $-0.83\pm0.05$\\
$\omega$ & [10$^{-14}$ s$^{-1}$] & $2.3\pm0.2$ & $2.7\pm0.4$\\
$\beta$ & [$10^{-3}$] & $\sim2.5$ & $\sim4.0$\\
$L/M$ & [$10^{-3}$ km s$^{-1}$ pc] & $1.0\pm0.2$ & $1.2\pm0.4$\\
\hline
\end{tabular}
\end{center}
\label{tab:grad}
\end{table}%

\section{Evaluation of new spectroscopic parameters for C$_{2}$H}

The C$_2$H radical is a linear molecule. Its unpaired electron causes a splitting of 
each rotational level into two fine structure levels. In addition, the spin of the 
H nucleus causes each fine structure level to be split further into two hyperfine levels. 
The strong transitions are those having $\Delta N = \Delta J = \Delta F$ 
which leads to four strong transitions at higher quantum numbers. 
However, transitions with $\Delta N \neq \Delta J$ or with $\Delta J \neq \Delta F$, 
which have essentially vanishing intensities at high $N$, 
may have fairly large relative intensities at lower values of $N$; 
the only restriction which has to be followed strictly is $\Delta F = 0$ or $\pm1$. 
In the case of the $N$=1$-$0 and $N$=2$-$1 transitions, this leads to 6 and 11 hyperfine 
lines, respectively, with non-zero intensity.


\subsection{Rest frequencies}
\label{restfreq}

Previous rotational data of \cch\ have been summarised by \citet{mk00}. 
They had measured submillimetre transitions up to 1~THz; the small hydrogen hyperfine splitting 
was not resolved. Their data set also included 6 hyperfine lines of the $N$=1$-$0 
measured toward a position north of Orion-KL \citep{gg83}, four hyperfine components 
of the $N$=2$-$1 transition obtained in the laboratory by the same authors 
as well as four $N$=3$-$2 hyperfine features \citep{sh81}. 
These data as well as predictions are available from  the recommended 
CDMS\footnote{\tt http://www.astro.uni-koeln.de/cdms/} 
(Cologne Database for Molecular Spectroscopy, M\"uller et al.~2001, 2005).

In the present investigation we have observed all six HFS components of the $N$=1$-$0 transition 
as well as seven components of the $N$=2$-$1 transition. Because of good to very good signal-to-noise 
(S/N) ratios of all these lines and because of the small line width it seemed promising 
to determine improved rest frequencies for these lines with respect to those available 
in the CDMS. 
The main sources affecting the accuracy of rest frequencies, besides S/N and line width 
are expected to be the symmetry of the (usually Gaussian) line shapes and the accuracy 
with which the velocity structure of the source is known.
On the basis of the well determined rest frequencies of \nnh(1$-$0) \citep{pd09} 
and \co(2$-$1) \citep{CDMS_1} we have derived at the dust continuum peak an LSR velocity  of 7.80~km~s$^{-1}$ for L1498 
based on the observations of these two transitions by \citet{tm02} 
and an LSR velocity of $-0.83$~km~s$^{-1}$ for CB246 from the present observations 
of these transitions (see Sect. \ref{nnh_coCB}). 
Employing these LSR velocities, rest frequencies of \cch\ have been determined separately 
for the two observed sources. They differ on average by 3 and 17~kHz, respectively, 
for the $N$=1$-$0 and $N$=2$-$1 transition and the averages for each HFS component are given in 
Table \ref{tab:c2h10p} and \ref{tab:c2h21p}, respectively together with the assignments, 
the estimated uncertainties, the residuals o$-$c between the observed frequencies 
and those calculated from the final set of spectroscopic parameters, 
and the relative intensities $f$.

\begin{table}[!ht]
\caption{Observed hyperfine structure components of the $N'\rightarrow N=1\rightarrow0$ transition in \cch: 
  rest frequencies$^a$, residuals o$-$c between observed and calculated frequencies$^b$, 
  and relative intensities $f$.} 
\label{tab:c2h10p} 
\centering 
\begin{tabular}{c c c r c} 
\hline\hline 
comp. & Transition & $\nu$ & o$-$c &  $f$ \\ 
no. & $J',F'\rightarrow J,F$ & [MHz] & [kHz] & \\  
\hline 
1 & $3/2,1\rightarrow1/2,1$ & 87284.105\,(10) & $-$3 & 0.042\\
2 & $3/2,2\rightarrow1/2,1$ & 87316.898\,(10) &    4 & 0.416\\
3 & $3/2,1\rightarrow1/2,0$ & 87328.585\,(10) & $-$2 & 0.207\\
4 & $1/2,1\rightarrow1/2,1$ & 87401.989\,(10) & $-$3 & 0.208\\
5 & $1/2,0\rightarrow1/2,1$ & 87407.165\,(10) &    8 & 0.084\\
6 & $1/2,1\rightarrow1/2,0$ & 87446.470\,(10) &    0 & 0.043\\
\hline 
\end{tabular}\\[2pt]
$^a$ Numbers in parentheses are one standard deviation in units of the least significant figures.\\
$^b$ Calculated from present set of spectroscopic parameters in Table~\ref{CCH_parameters}.
\end{table} 

\begin{table}[!ht]
\caption{Observed hyperfine structure components of the $N'\rightarrow N=2\rightarrow1$ transition 
  in \cch: rest frequencies$^a$, residuals o$-$c between observed and calculated frequencies$^b$, 
  and relative intensities $f$.} 
\label{tab:c2h21p} 
\centering 
\begin{tabular}{c c c r c} 
\hline\hline 
comp. & Transition & $\nu$ & o$-$c & $f$ \\ 
no. &$J',F'\rightarrow J,F$ & [MHz] & [kHz] & \\  
\hline 
1 & $5/2,2\rightarrow3/2,2$ & 174634.861\,(16) &     6 & 0.017\\
2 & $5/2,3\rightarrow3/2,2$ & 174663.199\,(12) &     0 & 0.350\\
3 & $5/2,2\rightarrow3/2,1$ & 174667.629\,(14) & $-$12 & 0.232\\
4 & $3/2,2\rightarrow1/2,1$ & 174721.744\,(14) &  $-$8 & 0.197\\
5 & $3/2,1\rightarrow1/2,0$ & 174728.071\,(12) &    12 & 0.083\\
6 & $3/2,1\rightarrow1/2,1$ & 174733.210\,(14) & $-$15 & 0.052\\
7 & $3/2,2\rightarrow3/2,2$ & 174806.843\,(14) &  $-$7 & 0.045\\
\hline 
\end{tabular}\\[2pt]
$^a$ Numbers in parentheses are one standard deviation in units of the least significant figures.\\ 
$^b$ Calculated from present set of spectroscopic parameters in Table~\ref{CCH_parameters}.
\end{table} 

\subsection{Spectroscopic parameters}
\label{discussion_parameters}

These rest frequencies, together with previous laboratory values \citep{sh81,mk00}, 
were used to calculate spectroscopic parameters for C$_2$H. As in \citet{mk00}, 
the rotational constant $B$, the quartic and sextic distortion terms $D$ and $H$, 
the electron spin-rotation parameter $\gamma$ along with its distortion correction 
$\gamma ^D$, as well as the scalar and tensorial electron spin-nuclear spin coupling terms 
$b_F$ and $c$ were determined. The distortion correction $b_F^D$ to $b_F$ was not 
determined with significance, even though its absolute uncertainty was slightly smaller than 
in \citet{mk00}, and it did not contribute significantly to the reduction of the 
rms error. It was thus omitted from the final fit. However, it was found that the 
$^1$H nuclear spin-rotation constant $C$ improved the quality of the fit 
by a non-negligible amount of 12\,\% even though it was barely determined. 
As, in addition, its value appeared to be reasonable, see 
further below, this constant was retained in the final fit. 
The resulting spectroscopic parameters are given in Table~\ref{CCH_parameters} 
together with the most recent previous values by \citet{mk00}.


\begin{table}[!ht]
\caption{Spectroscopic parameters$^a$ (MHz) of C$_2$H in comparison to previous values.}
\label{CCH_parameters}
\centering
\begin{tabular}{l r@{}l r@{}l}
\hline\hline
Parameter & \multicolumn{2}{c}{present value} & \multicolumn{2}{c}{\citet{mk00}} \\
\hline
$B$                     & 43674&.5177\,(13) & 43674&.5289\,(12)  \\
$D \times 10^3$         &   105&.515\,(53)  &   105&.687\,(51)   \\
$H \times 10^6$         &  $-$0&.42\,(32)   &     0&.32\,(32)    \\
$\gamma$                & $-$62&.6103\,(45) & $-$62&.6029\,(43)  \\
$\gamma ^D \times 10^3$ &  $-$2&.227\,(250) &  $-$2&.313\,(255)  \\
$b_F$                   &    44&.4788\,(82) &    44&.4922\,(183) \\ 
$b_F^D$                 &      &$-$         &  $-$0&.0100\,(38)  \\
$c$                     &    12&.2389\,(232)&    12&.2256\,(261) \\
$C$                     &  $-$0&.0087\,(53) &      &$-$          \\
\hline
\end{tabular}\\[2pt]
$^a$ Numbers in parentheses are one standard deviation in units of the least significant figures. 
\end{table} 

The most striking feature in the comparison of the present C$_2$H rest frequencies with 
those from \citet{gg83} is that the latter are on average 28.6~kHz higher in frequency, 
ranging from complete agreement to 51~kHz for the individual HFS components. 
Moreover, the deviations are a few times the uncertainties reported by \citet{gg83} for 
some lines. In fact, employing the present spectroscopic parameters given in 
Table~\ref{CCH_parameters}, the \citet{gg83} data are reproduced to only 3.6 times 
their reported uncertainties on average. We suspect that the LSR velocity of the source 
north of Orion-KL used in \citet{gg83} was not as well known as the authors assumed 
and that for the laboratory measurements the frequency determinations were slightly off 
or the estimates of the uncertainties were slightly too optimistic.

The present rest frequencies have been reproduced to 0.6 times the uncertainties, 
slightly better still for the $N$=1$-$0 transition, suggesting that the uncertainties 
in Table~\ref{tab:c2h21p} and even more so in Table~\ref{tab:c2h10p} have been judged 
somewhat too conservatively. On the other hand, uncertainties in the LSR velocity 
may justify such a conservative error estimate.
The $N$=3$-$2 lines from \citet{sh81} have been reproduced to better than 18~kHz, 
which is much better than the $\sim 45$~kHz with which these data where reproduced 
in \citet{mk00}. In other words, these $N$=3$-$2 rest frequencies are 
much better compatible with the present $N$=1$-$0 and $N$=2$-$1 rest frequencies 
than with those from \citet{gg83}. The submillimetre data from \citet{mk00} are 
at somewhat higher frequencies and quantum numbers such that both present and 
previous spectroscopic parameters in Table~\ref{CCH_parameters} reproduce these data 
on average to about 0.6 times the reported uncertainties.

While frequency deviations of a few tens of kHz may possibly be neglected in 
observations of hot cores or similar sources, the deviations are rather considerable 
for investigations into the dynamics of dark clouds as in the present study. 
The deviations are also reflected in the small ($\sim 11$~kHz) differences in the 
rotational constant $B$, see Table~\ref{CCH_parameters}. Nevertheless, these 
deviations are more than four times the combined uncertainties and thus clearly significant. 
It is worth noting that the $B$ value in \citet{mk00} is determined essentially by 
the data from \citet{gg83}. The slightly higher $D$ value from the present study, 
just outside the combined uncertainties, compensates the change in $B$ to some degree 
for the submillimetre lines. The sextic term $H$ is essentially the same as in 
the previous study, still not determined with significance, but probably of the 
right order of magnitude.

Changes in the fine structure parameters and in the larger hyperfine structure 
parameters are essentially insignificant. The distortion correction $b_F^D$ 
had been used in the previous fit of \citet{mk00} since its inclusion improved the 
\citet{gg83} data to be reproduced on average from 1.31 times the uncertainties 
to 0.93 times the uncertainties. In present trial fits its value was determined 
as $-0.0021\,(33)$~MHz. The magnitude of the ratio $b_F^D/b_F$ is now much closer 
to that of $\gamma ^D/\gamma$, but still much bigger than that of $D/B$. 
However, the uncertainty was larger in magnitude than the value, and the inclusion of 
the parameter in the fit improved the quality of the fit only insignificantly. 
The term $b_F^D$ was consequently omitted from the final fit.
The inclusion of the $^1$H nuclear spin-rotation term $C$ in the present fit 
requires some explanation. This term is usually small and negative and frequently 
scaling with the rotational constant is the dominant contribution to its size.
The C$_2$H value of $-$8.7\,(53)~kHz agrees within its large uncertainty 
with the experimental one of $-$4.35\,(5)~kHz for HCN \citep{HCN_dipole_HFS} 
and with the calculated values of $-$4.80~kHz and $-$5.55~kHz for HCN and 
HCO$^+$ \citep{H13CO+_HFS} and appears thus to be reasonable as these species have 
fairly similar rotational constants.


\section{Non-LTE hyperfine populations}



In this Section we consider the evidence for non-LTE populations in the 
hyperfine levels 
sampled by our observations. 
We first compare the observed intensity ratios of 
\cch\, 
in L1498 and CB246 with the predictions of simple LTE models and find 
that while real deviations 
from LTE populations are present, the LTE assumption is an approximation 
which is useful for many purposes. 
We then consider a Monte Carlo radiative transfer program for the case of 
\cch\, 
and show that the observed deviations can be approximately understood 
with an educated guess 
at the (unknown) \cch\, collisional rates.

\subsection{Single- and two-layer models}

\begin{figure}[!h] 
\resizebox{\hsize}{!}{\includegraphics[angle=0]{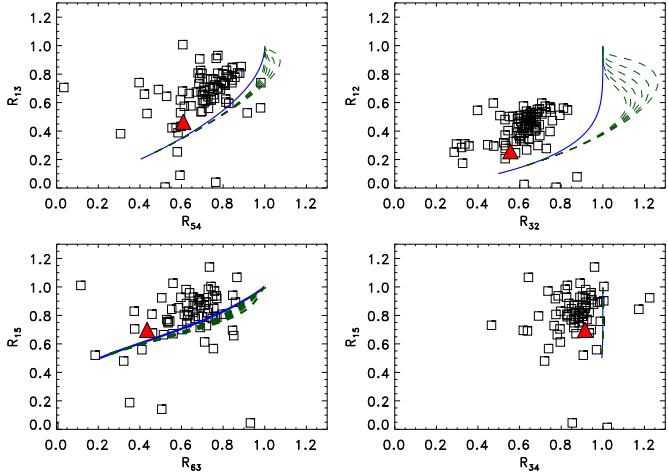}} 
\caption{Ratio of the integrated intensities of some couples of components of \cch(1$-$0) in L1498, where $R_{ij}$ represents the ratio between $\int T_{{\rm mb},i}{\rm d}V$ and $\int T_{{\rm mb},j}{\rm d}V$. Observational data, {\em black empty squares}; one-layer model, {\em blue solid curve}; two-layer model with different $T_{{\rm ex},f}$, {\em green dashed curves}; results from Monte Carlo model, {\em red filled triangles}. The typical errors on the ratios are of the order of $8.5\times10^{-2}$.} 
\label{area_nnvsnnL_idl_MC_last} 
\end{figure}

\begin{figure}[!h] 
\resizebox{\hsize}{!}{\includegraphics[angle=0]{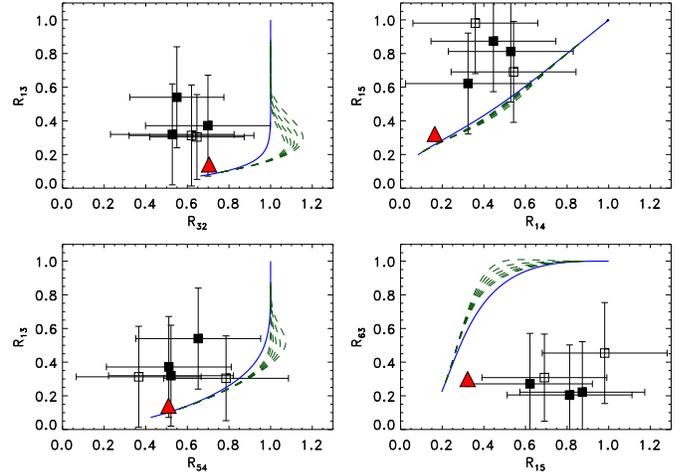}} 
\caption{Ratio of the integrated intensities of some couples of components of \cch(2$-$1), where $R_{ij}$ represents the ratio between $\int T_{{\rm mb},i}{\rm d}V$ and $\int T_{{\rm mb},j}{\rm d}V$. Observational data, {\em black empty squares} for L1498 and {\em black filled squares} for CB246; one-layer model, {\em blue solid curve}; two-layer model with different $T_{{\rm ex},f}$, {\em green dashed curves}; results from Monte Carlo model for L1498, {\rm red filled triangles}.} 
\label{area_nnvsnnLCB_21} 
\end{figure}

We note first that for the case of LTE between different hyperfine levels 
(i.e. all transitions of a given multiplet have the same excitation 
temperature), 
we expect that for a homogeneous slab, the ratio of line intensities of 
two transitions $R_{ij}$ is given by:

\begin{equation}
R_{ij}=\frac{1-\exp(-f_{i}\tau)}{1-\exp(-f_{j}\tau)}
\label{Rij}
\end{equation}
where $\tau$ is the total transition optical depth and $f_{i}$ is the 
relative line strength of the $i$-th component as e.g. in Table 
\ref{tab:c2h10p}.
Thus, as the optical depth varies, the expected $R_{ij}$ varies from 
$f_{i}/f_{j}$ to unity and for two different pairs of transitions, one 
can derive a curve in the plane ($R_{ij},R_{kl}$). 
In Fig. \ref{area_nnvsnnL_idl_MC_last}, we give for various combinations 
of transitions in the $N$=1$-$0 transition our observed line ratios 
compared with the result expected on the basis of equation \ref{Rij}, and 
in Fig. \ref{area_nnvsnnLCB_21}, we show analogous results for the 
$N$=2$-$1 
line. We see that the observed intensity ratios in \cch(1$-$0) are not 
consistent with equation \ref{Rij} although the deviations are not large. 
In particular, we note in Fig. \ref{area_nnvsnnL_idl_MC_last} that ratios 
such as $R_{32}$ (upper right panel) at some positions attain values 
below the optically thin expected value of 0.5. More significantly 
perhaps the observed ratio $R_{34}$ (bottom right panel) of the 83728 and 
87401 MHz lines which have essentially the same line strength is 
typically of order 0.8$-$0.9 and at almost all positions less than unity. 
These differences are not significant at individual positions, but appear 
to be significant statistically. The results in Fig. 
\ref{area_nnvsnnL_idl_MC_last} are for L1498, but similar trends hold for 
CB246.

Fig. \ref{area_nnvsnnLCB_21} shows analogous results for \cch(2$-$1). 
There are fewer positions but again, we see significant deviations from 
the homogeneous single layer prediction of equation \ref{Rij}. The ratio 
$R_{32}$ (upper left panel) of the 174667 and 174663 MHz lines is 
typically between 0.5 and 0.6 as compared to the optically thin LTE 
prediction of 0.66. These differences are small but appear to be 
significant.


Equation \ref{Rij} assumes a homogeneous one-dimensional solution of the transfer equation and it is clear that for a real prestellar core, gradients in both temperature and density are present. Indeed observations of some cores in strong ground state transitions such as HCO$^{+}$(1$-$0) are complicated by absorption in a foreground layer whose excitation appears to be essentially that of the cosmic background (Tafalla et al.~1998).
In the particular case of \cch\, towards L1498, the ``self-absorption'' observed in the 87316 MHz line (see Fig. \ref{c2h_2007obs}) suggests that something of this sort may occur also for \cch\,. In view of this, we have also considered a two-layer model of the type discussed by Myers et al.~(1996) with background ($b$) and foreground ($f$) layers having excitation temperatures $T_{{\rm ex},b}$ and $T_{{\rm ex},f}$ and optical depths $\tau_{f}$ and $\tau_{b}$ respectively. We then find for the emergent intensity of $i$-th component:



\begin{eqnarray}
T_{{\rm mb},i} & = & J(T_{{\rm ex}, f})[1-e^{-\tau_{f,i}}]+J(T_{{\rm ex},b})[1-e^{-\tau_{b,i}}]e^{-\tau_{f,i}}\nonumber\\
& &-J(T_{\rm bb})[1-e^{-\tau_{b,i}-\tau_{f,i}}]
\label{eq:2layer}
\end{eqnarray}
where

\begin{equation}
J(T)=\frac{T_{0}}{\exp(T_{0}/T)-1}
\label{eq:jt}
\end{equation}
is the Planck-corrected brightness temperature and $T_{\rm bb}=2.73$ K 
is the temperature of the cosmic background. 
$T_{0}\equiv h\nu/k$, where $\nu$ is the transition frequency and $h$ 
and $k$ represent Planck's and Boltzmann's constants respectively. 
The optical depth of the $k$-layer ($k=f,b$) for the $i$-the component is defined 
as $\tau_{k,i}\equiv\tau_{0}f_{i}\phi_{i}$, where $f_{i}$ is again 
the relative intensity and $\phi_{i}$ is the profile which we assume as gaussian, 
that is

\begin{equation}
\phi_{i}=2\sqrt{\frac{\ln2}{\pi}}\exp\left\{-\frac{4[V-(-1)^{m}\langle V_{k,i}\rangle]^{2}\ln2}{\sigma^{2}}\right\}
\end{equation}
with $m=0$ for the foreground layer and $m=1$ for the background layer. We have included in Fig. \ref{area_nnvsnnL_idl_MC_last} and \ref{area_nnvsnnLCB_21} predictions based on our two-layer model varying $T_{{\rm ex},f}$ and we derive the dashed curves using equation \ref{eq:2layer}. We see from this that the effect of ``foreground layers'' is to increase ratios such as $R_{32}$ in some cases to values of
order unity. Essentially, this is due to absorption of the strong 87316 MHz line (component 2). One sees from both Fig. \ref{area_nnvsnnL_idl_MC_last} and \ref{area_nnvsnnLCB_21} that this is not what is required to fit the observed data and in fact the two-layer models are a poorer approximation to the observed intensity ratios than the single-layer LTE model. We thus conclude that understanding the observed line ratios requires a proper non-LTE treatment.


\subsection{Monte Carlo treatment of radiative transfer in the
C$_2$H rotational transitions}
\label{sec:MC}

The two-layer model, with its 
assumption of the same excitation temperature for all 
hyperfine components in each layer and its
discontinuous jump in conditions between the two emitting regions,
misses an important 
part of the complex pattern of level populations responsible for
the emission in \cch(1$-$0). 
As can be seen in Fig. \ref{c2h_2007obs}, the hyperfine components
of C$_2$H(1$-$0) are simultaneously subthermal and optically thick 
in both L1498 and CB246, and under these conditions, the excitation
temperature of each component is determined by a delicate
balance between collisions and trapping. This balance will be different
for each transition depending on its particular optical depth,
and it will therefore give rise to differences in excitation
between the components of the $N$=1$-$0 multiplet.
The large optical depth of the lines, in addition, makes the emission
from each component originate in gas at different depth in the core,
and this also contributes to differences in the observed line intensities.
Such complex interplay between excitation and 
optical depth effects cannot be treated accurately with the two-layer
model, and its analysis requires a more sophisticated numerical scheme.
In this Section, we present the result of a (partial) solution to
the C$_2$H radiative transfer problem using the same Monte Carlo code
used in Tafalla et al.~(2004, 2006) to analyse the emission from
a number of molecular species in the L1498 and L1517B cores. 

As the modelling of the L1498 emission in Tafalla et al.~(2004, 2006) 
already fixed the physical description of this
core (radial profiles of density, temperature, and
velocity), the only parameter left free to model 
the C$_2$H emission is the radial profile of abundance. 
Unfortunately, the C$_2$H analysis 
is limited due to the lack of known collision rates for the 
species, and this forces us to make an educated guess of
this important set of coefficients. Following Turner et al.~(1999),
we approximate the collision rates of C$_2$H using
those for HCN calculated by Green \& Thaddeus (1974). From these rates,
which do not include hyperfine structure, we derive a new
set of rates with hyperfine structure assuming that the
new rates are simply proportional to the degeneracy of the 
final state (Guilloteau \& Baudry 1981, Lique et al.~2009). 
Additional $\Delta N =0$ rates were included by following the recipe from 
Turner et al.~(1999), and
a total of 30 levels with 37 transitions (up to $N=7$ 
and an energy equivalent to 120 K) were used in the calculation.

\begin{figure}
\centering
\resizebox{\hsize}{!}{\includegraphics{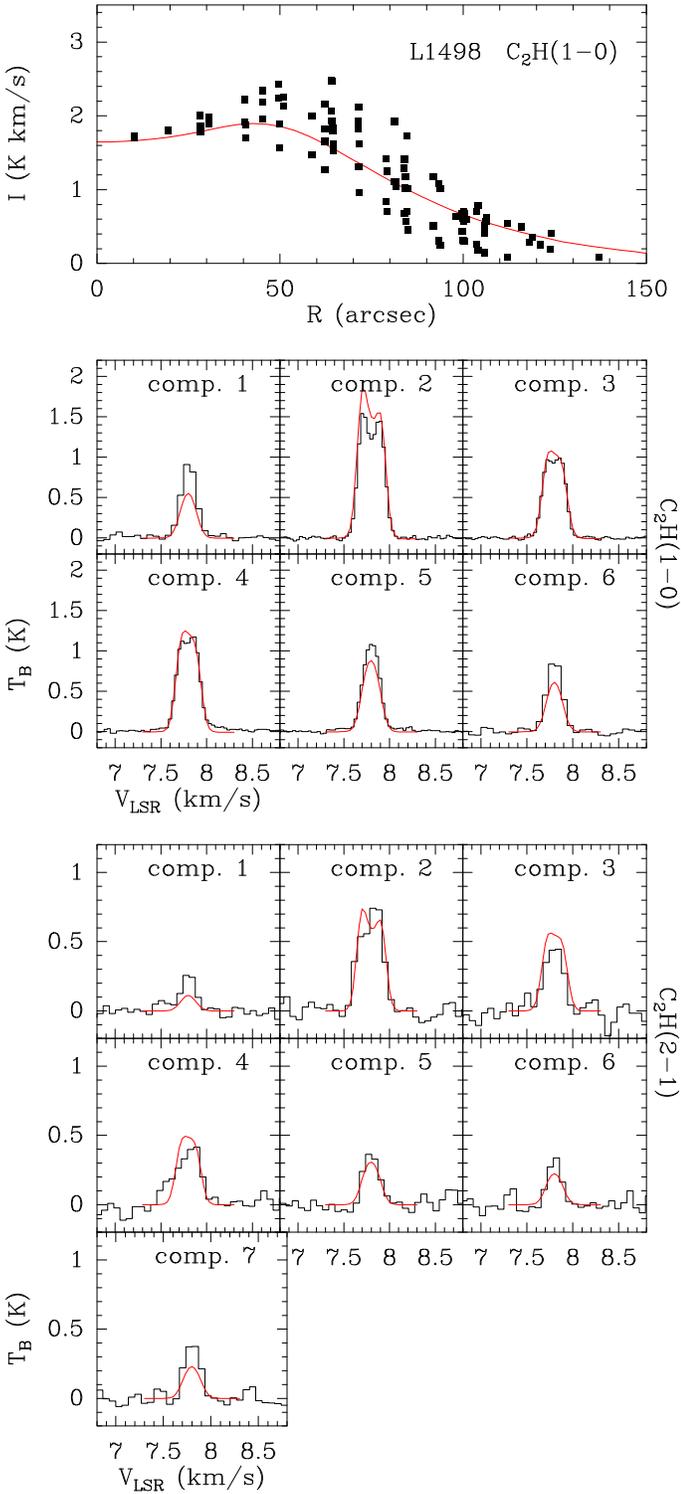}}
\caption{Comparison between observations and our best fit Monte Carlo 
model of the C$_2$H emission in L1498. 
{\em Top panel: } radial profile of observed C$_2$H(1$-$0) intensity
integrated over all hyperfine components ({\em filled squares}) and
model prediction for a constant abundance core with a central 
depletion hole ({\em solid red line}).
{\em Middle panels: } emerging spectra for each component of the
C$_2$H(1$-$0) multiplet ({\em black histograms}) and predictions
for the same best fit model ({\em solid red lines}).
{\em Bottom panels: } same as middle panels but for the components
of the C$_2$H(2$-$1) multiplet.
Note the reasonably good fit of all observables 
despite the use of highly approximated collisional rates (see text).
\label{fig_mc_fits}}
\end{figure}

Following the analysis of other species in L1498,
the goal of our C$_2$H modeling was to fit simultaneously
the combined radial profile of $N=$1$-$0 integrated intensity
together with the central spectra of the different components of
the $N$=1$-$0 and $N$=2$-$1 multiplets, assuming a uniform kinetic
temperature of 10 K (as derived from a non-LTE analysis of the
NH$_{3}$ data by Tafalla et al. 2004).  A first set of model runs
using the collision rates described before 
predicted excitation temperatures that were too high
in the outer core layers, a situation that is 
inconsistent with the self absorbed profile seen in the
thickest $N=$1$-$0 component (Fig. \ref{c2h_2007obs}).
This inconsistency indicated that the guessed C$_2$H
collision rates were too large, and that they should
be significantly reduced in order to match the observations.
To avoid introducing artifacts in the relative excitation
of the hyperfine components, the collision rates were reduced
dividing 
them by a global factor of 3. With these corrected collision rates,
a fit was achieved by assuming a constant C$_2$H abundance 
with respect to H$_{2}$
of $8 \times 10^{-9}$ and a central depletion region of radius
$9 \times 10^{16}$~cm inside which the \cch\ abundance is
negligible (10$^{-4}$ times the outer value), 
similar to that of other species
in L1498, see Tafalla et al.~2006. The results of this model
are shown in Fig.~\ref{fig_mc_fits}  (red lines) superposed to 
observations (black
histograms and squares). As it can be seen, the model fits reasonably well
both the radial profile of $N$=1$-$0 intensity and the spectra of
most components in both the $N$=1$-$0 and $N$=2$-$1 multiplets. 
The model in addition, fits reasonably well
the observed ratios of line pairs presented in Fig. \ref{area_nnvsnnL_idl_MC_last}.

\begin{figure}
\centering
\resizebox{\hsize}{!}{\includegraphics{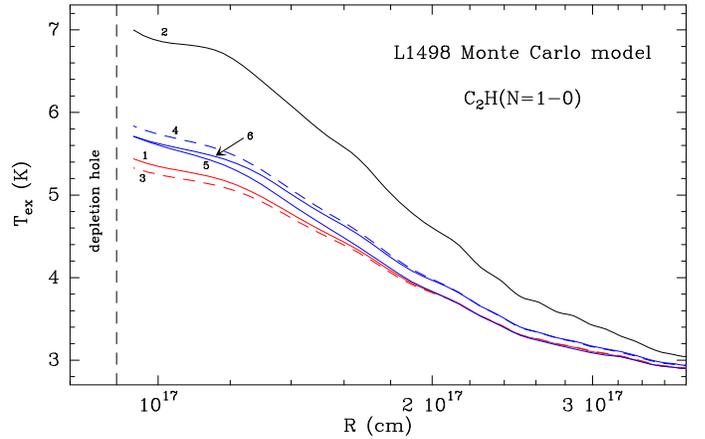}}
\caption{Radial profile of excitation temperature for each hyperfine
component of C$_2$H(1--0) as predicted by our best fit Monte
Carlo model (each component is labelled according to the
ordering in Table 3). Note the gradual drop of $T_{\mathrm ex}$ with
radius, which is caused by the combined decrease in collisional
excitation and photon trapping towards the outer core.
The different hyperfine components have different
$T_{\mathrm ex}$ depending mostly on trapping effects
(see text for a full discussion).
\label{fig_mc_tex}}
\end{figure}

If our Monte Carlo model fits the 
observed C$_2$H emission, we can 
use it to analyze the excitation of the different 
$N$=1$-$0 components and understand
the origin of the observed line ratios. This is best done
by studying the radial profile of excitation temperature,
which is presented in Fig.~\ref{fig_mc_tex}.
As the figure shows, the $T_{\rm ex}$ of each component 
systematically decreases with radius,
from about 6~K in the core interior to a value close to
the cosmic background temperature
near the core edge. The figure, in addition, shows that
at each radius, the $T_{\rm ex}$ of the different components can differ
by as much as 1~K, and that component 2 is significantly
more excited than the rest. These differences in excitation
over the core and among components
could in principle result from differences in the
contribution of collisions or from photon trapping, and
to disentangle the two effects 
we have run an alternative model having
a factor of $10^4$ lower abundance. This optically 
thin case also presents an outward drop in $T_{\rm ex}$,
this time entirely due to the effect of collisions, 
but it predicts excitation temperatures in the core interior that
are about 1~K lower than for the best fit model. 
In addition,
component 2 in this thin case has a 
$T_{\rm ex}$ comparable to that of the other components,
and the overall scatter of $T_{\rm ex}$ 
among all components does not exceed
0.5~K. This decrease in the excitation 
when the lines become thin indicates that
in the best fit model
the $N=1$ level populations are significantly enhanced by trapping
of $N=$1$-$0 photons. The higher excitation of component 2,
in particular, appears as an extreme case of trapping:
this component has the largest relative intensity
and therefore is the most sensitive one to optical depth effects.
Its sensitivity to trapping makes it
the brightest line of the multiplet despite 
suffering from self absorption at the line centre.
Differential line trapping seems also responsible
for the different intensity of components 3 and 4, which would
otherwise be equally bright
because of their equal Einstein $A$ coefficient and upper
level statistical weight. As Fig. \ref{fig_mc_tex} shows, component 4 
has an $\approx 0.5$~K higher $T_{\rm ex}$ in the
inner core, and this is most likely the result of enhanced
trapping due to an overpopulation of the $N,J,F=0,1/2,1$
level, where most $N$=1$-$0 transitions end (except for
components 3 and 6).

In summary, our model shows that most excitation ``anomalies''
of the hyperfine components in the $N=$1$-$0 multiplet 
can be explained
as resulting from the different balance between collisions and trapping
expected for lines of very different intrinsic intensities under
conditions of subthermal excitation and high optical depth.
Further work on this issue requires an improved set of collision
rates for C$_2$H, and we encourage collision rate 
modelers to consider this species
for future computations.


\section{Column density and abundance estimate}
 Deriving a column density for C$_{2}$H requires an estimate of the excitation temperature. This
 as we see in the Monte Carlo calculations discussed above depends on
 the line transfer and hence on position within the core. However a
 reasonable approximation to the column density can be obtained assuming
 a homogeneous layer with  constant excitation temperature. We have
 verified this assumption for the case of L1498  using the Monte Carlo program.

 The excitation temperature $T_{\rm ex}$
 can be inferred ({\em method 1}) from an LTE fit to the hyperfine
 components of either the $N$=1$-$0 or $N$=2$-$1 lines assuming unity beam filling factor
 and using the measured intensity of optically thick transitions.  An independent
 measure can be obtained ({\em method 2}) from the ratio of intensities of low line strength
 transitions of the $N$=2$-$1 and $N$=1$-$0 lines assuming them to be optically thin.
 We summarize in Table \ref{tab:Tex} the excitation temperatures derived 
 using these different approaches for
 the positions where we have $N$=2$-$1 data available. From Table \ref{tab:Tex} we see that
 in both sources, the excitation temperature appears to be between 3.8 and 4.9
 K. We will assume an excitation temperature of 4 K in the following for
 all positions.

\begin{table}[!h]
\caption{Values of $T_{\rm ex}$ [K] derived using the two  
methods.}
\begin{center}
\begin{tabular}{cccc}
\hline
$\alpha,\delta$ offsets&\multicolumn{2}{c}{\em method 1}&{\em method 2}\\
$['',''$] & C$_{2}$H(1$-$0) & C$_{2}$H(2$-$1) & \\
\hline
\multicolumn{4}{c}{L1498}\\
0,0 & $4.0\pm0.1$ & $4.6\pm0.6$ & $4.2\pm0.6$\\
60,$-$40 & $4.1\pm0.1$ & $4.3\pm0.8$ & $4.4\pm0.5$\\
\hline
\multicolumn{4}{c}{CB246}\\
0,0 & $4.1\pm0.1$ & $4.9\pm1.0$ & $4.4\pm0.9$\\
15,$-$30 & $4.1\pm0.3$ & $4.4\pm0.7$ & $4.4\pm0.4$\\
--60,15 & $4.0\pm0.2$ & $3.9\pm0.4$ & $3.8\pm0.6$\\
\hline
\end{tabular}
\end{center}
\label{tab:Tex}
\end{table}%

For an optically thin  \cch(1$-$0) line, one can
 derive the column density, $N({\rm C_{2}H})$, through the formula:

\begin{equation}
N({\rm C_{2}H})=\frac{8\pi\nu_{ji}^3}{c^3}\frac{Q}{A_{ji}g_j}\frac{e^{E_j/kT_{ex}}}{e^{h\nu_{ji}/kT_{ex}}-1}\frac{\int T_{\rm mb}\mathrm{d}v}{J(T_{ex})-J(T_{bg})}
\label{IItoNC2H}
\end{equation}
where $\nu_{ji}$ is the transition frequency, $c$ the speed of light, $A_{ji}$ the
Einstein coefficient, $g_{j}$ the statistical weight,
$E_{j}$ the energy of the upper level, $T_{\rm ex}$ the excitation temperature,
$\int T_{\rm mb}\mathrm{d}v$ the integrated intensity, $J(T)$ as defined
in eq. \ref{eq:jt} and $Q=\sum_{j=0}^{\infty}g_{j}e^{-E_{j}/kT_{\rm ex}}$ 
is the partition function. The integral in equation \ref{IItoNC2H} 
is over all six hyperfine components.
In practise,  the stronger components of the line are sometimes thick and we
have hence integrated only over the 87284 MHz and 87446 MHz lines (components 1
and 6 in Table \ref{tab:c2h10p}) and divided the result by the sum of their line strengths
(0.085) in order to evaluate the integral in equation \ref{IItoNC2H}. 


 We need to compare this with the molecular hydrogen column density $N$(H$_{2}$)
 and  we do this using the mm dust emission (from  Tafalla et al.~2004 for L1498 
and from this work, see Sect. \ref{boloCB}, for CB246).   
We assume here in both sources a dust grain opacity of
 0.005 cm$^{2}$ g$^{-1}$ of dust and a dust temperature of 10 K, as in Sect. \ref{MassCB246}.

We also evaluated the column densities and the abundances of \nnh and CO 
for CB246 (using our data, see Sect. \ref{nnh_coCB}) and for L1498 
(using data from Tafalla et al.~2004). We derived $N$(\nnh) following 
the same procedure as for $N$(\cch). Thus, if the stronger components 
of \nnh(1$-$0) are thick then we integrated only over the isolated component 
($F'_{1},F'\rightarrow F_{1},F=1,2\rightarrow0,1$), dividing the result by 
its line strength (0.111). In the case of \co, we used directly equation 
\ref{IItoNC2H} to calculate $N$(\co) and $N$(CO) assuming [$^{16}$O]/[$^{18}$O]$\sim$560 
(Wilson \& Rood 1994).  We give in Table \ref{tab:abu} abundance estimates 
that we have made at selected positions in L1498 and CB246.

\begin{table}[!h]
\caption{Molecular abundances in L1498 and CB246.}
\begin{center}
\begin{tabular}{cccc}
\hline
$\alpha,\delta$ offsets & [C$_{2}$H]/[H$_{2}$] & [N$_{2}$H$^{+}$]/[H$_{2}$] & [CO]/[H$_{2}$]\\
$[$$'',''$] & [10$^{-8}$] & [10$^{-10}$] & [10$^{-5}$]\\
\hline
\multicolumn{4}{c}{L1498}\\
$0,0$ & $0.8\pm0.1$ & $1.9\pm0.4$ & $0.5\pm0.1$\\
$-40,20$ & $1.1\pm0.3$ & $1.8\pm0.5$ & $0.8\pm0.2$\\
$60,-40$ & $1.5\pm0.4$ & $2.4\pm0.7$ & $1.1\pm0.3$\\
\hline
\multicolumn{4}{c}{CB246}\\
$0,-15$ & $1.0\pm0.3$ & $3.1\pm1.6$ & $0.9\pm0.2$\\ 
$-60,15$ & $1.2\pm0.5$ & $2.4\pm1.7$ & $1.0\pm0.3$\\
$-60,60$ & $1.0\pm0.3$ & $1.2\pm0.9$ & $1.3\pm0.4$\\
$-30,15$ & $1.0\pm0.2$ & $3.0\pm1.2$ & $1.6\pm0.5$\\ 
\hline
\end{tabular}
\end{center}
\label{tab:abu}
\end{table}%

\begin{figure}[!h]
\begin{center} 
\resizebox{\hsize}{!}{\includegraphics[angle=0]{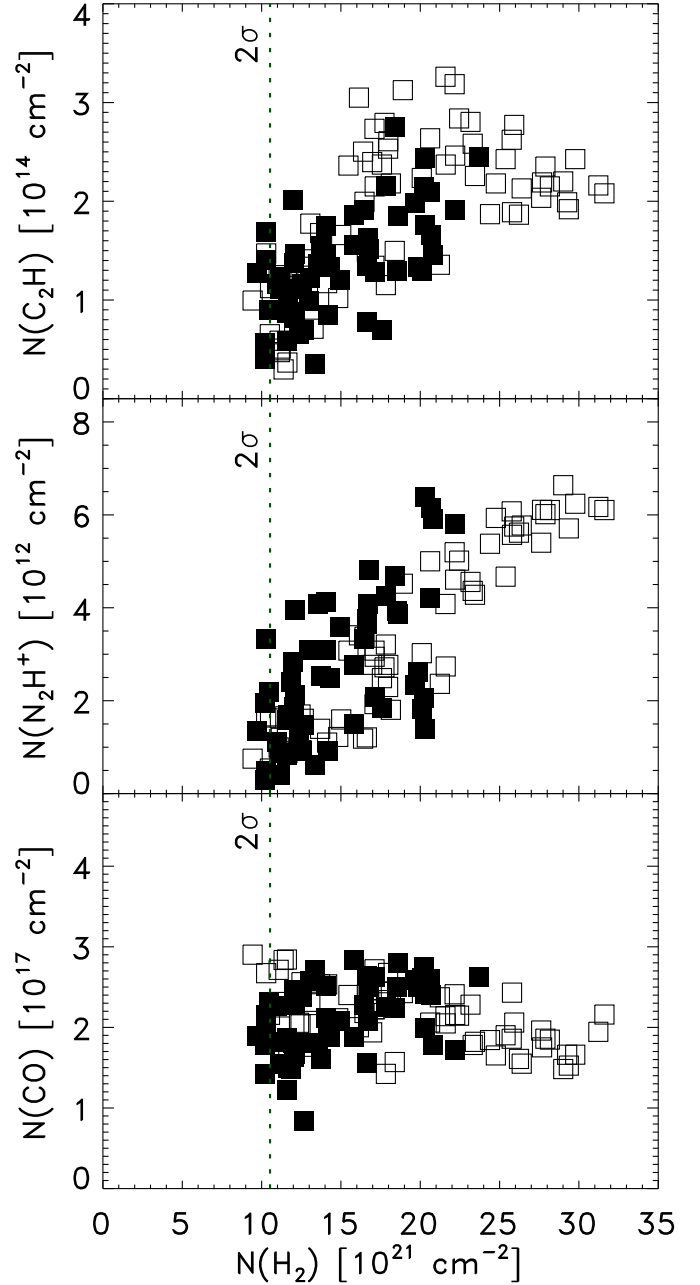}} 
\caption{Plot of the column density of \cch, \nnh\ and CO against the H$_{2}$ 
column density for L1498 ({\em open squares}) and CB246 ({\em filled squares}). 
The dotted line represents the $2\sigma=6$ mJy limit of the continuum emission. 
Data for \nnh\ and CO for L1498 are from Tafalla et al.~(2004).} 
\label{LCB_cont_vs_C2H10}
\end{center} 
\end{figure}

In Fig.~\ref{LCB_cont_vs_C2H10}, we show the plot of \cch, \nnh\ and CO 
column
density against  H$_{2}$ column density at different positions in both 
sources.
 We find that for column densities 
 $N($H$_{2}$)$\lesssim2\times10^{22}$ cm$^{-2}$, $N$(\cch) is 
 proportional to
 $N$(H$_{2}$) but, in L1498, for higher values of $N($H$_{2}$), this 
 proportionality breaks
 down and $N($\cch) seems to saturate at a value of $\sim2.2\times10^{14}$ cm$^{-2}$. This 
 is likely due to depletion of \cch\, in the high density core of L1498 as already
 suggested by the maps in Fig.~\ref{cont_C2H(2)L_crosses}, the cuts in Fig.~\ref{cut_paper} and the Monte Carlo modeling.
 This behaviour is in complete analogy with
 several other species including the carbon rich molecule C$_{3}$H$_{2}$ 
 (Tafalla et al.~2006).
 
 On the other hand, for $N$(H$_{2}$) less than $2\times 10^{22}$ cm$^{-2}$, we
 find that the ratio $N$(\cch)/$N$(H$_{2}$) is constant corresponding to a constant
 \cch\, abundance in the low density part of the core. This corresponds to
 an average estimated \cch\, abundance relative to H$_{2}$ of 
$(1.0\pm0.3)\times10^{-8}$ in L1498 and $(0.9\pm0.3)\times10^{-8}$ in
 CB246. Our \cch\, abundances estimate for L1498 using eq. \ref{IItoNC2H} 
agrees to within 20\% with the Monte-Carlo estimate discussed in Section \ref{sec:MC}. 
It is noteworthy and somewhat surprising to us that the \cch\, abundances are so similar 
in cores of differing characteristics.
 
A more strict correlation over the whole range of observed H$_{2}$ column density 
is shown by the ratio $N$(\nnh)/$N$(H$_{2}$) which corresponds to an average estimated 
\nnh\ abundance relative to H$_{2}$ of $(1.8\pm1.2)\times10^{-10}$ in L1498 
and $(1.8\pm0.8)\times10^{-10}$ in CB246. \co\ behaves differently showing 
a rather constant value of its column density over the $N$(H$_{2}$) range corresponding 
to $N$(CO)=$(2.0\pm0.4)\times10^{17}$ cm$^{-2}$ in CB246 and 
$(2.1\pm0.3)\times10^{17}$ cm$^{-2}$ in L1498, while the average abundance relative to 
H$_{2}$ is $(1.3\pm0.6)\times10^{-5}$ in L1498 and $(1.5\pm0.4)\times10^{-5}$ in CB246.

  One difference between CB246 and L1498 is that the $N(\mathrm{H}_{2})$ peak 
(see Fig. \ref{LCB_cont_vs_C2H10}) is a factor 1.33 smaller
  in CB246 and indeed does not reach values for which large depletion is
  noted in L1498. However, our map in \co(2$-$1) (see lower panel of Fig. \ref{CB_twomaps_paper}) as
  well as the cuts shown in Fig. \ref{cut_paper_appendix} suggest to us that some CO depletion
  does occur in CB246.

\section{Discussion} 
\label{discussion}

 In this study, we have compared two cores of rather different 
 characteristics. L1498 is embedded in the Taurus complex with a
 total dimension of $2.6\times2.0$ arc minutes (Caselli et al.~2002) corresponding to $0.10\times0.08$ parsec at the distance of 140 parsec (Elias, 1978). CB246 is  an isolated globule with dimensions
 of $7.8\times 4.5$ arc minutes corresponding to $ 0.45 \times 0.26$ parsec at our adopted distance of 200 parsec.  It is thus perhaps not
 surprising that the non-thermal component of the line width, obtained from the observed line widths corrected for thermal broadening (see e.g. Caselli \& Myers 1995), is
 0.23 km s$^{-1}$ in CB246 as compared to 0.14 km s$^{-1}$ in L1498. However, apart from that,
 many characteristics of these two objects seem similar as discussed below. 

\subsection{Non-LTE effects}
 One of our aims was to  test for non-LTE effects in \cch\, if any  in
 the hope that this could lead to a better understanding of
 physical conditions in cores. Our results  show that non-LTE
 effects do occur and they are qualatively similar in the two
 cores.  With our Monte Carlo radiative transfer program, we have
 shown that one can qualitatively understand their nature in terms of
 trapping of individual components of the $N$=1$-$0 and $N$=2$-$1 lines. These
 calculations also show that reasonable estimates of parameters such
 as the \cch\, column density can be obtained using a simple one-layer
 homogeneous model with constant excitation temperature.  Given
 that real cores are not spherically symmetric, this is a useful
 simplification.  Clearly, more refined models will eventually be
 useful but equally clearly, this requires computations of
 collisional rates for \cch\, along the lines of those carried out
 by Monteiro and Stutzki (1986) for HCN. 

\subsection{Chemistry}
There has been significant recent work on the chemistry of prestellar
cores (see e.g. Aikawa et al.~2003, 2005, Lee et al.~2004,
Flower et al.~2006, Akyilmaz et al.~2007).  These studies follow
the evolution of molecular abundances in a collapsing prestellar core
as a function of time and initial conditions. 
For example, the Aikawa et al.~(2005) model, 
a model with $\alpha$, the ratio of gravitational to pressure force, 
equal to 1.1, predicts a [\cch]/[H$_{2}$] abundance ratio of $4\times10^{-9}$ 
at a radius of 10000 AU (i.e. outside the depletion hole) or a factor of 
roughly 3 smaller than our observed value but there are a variety of factors 
which influence this estimate as we now discuss. One factor is that results 
are sensitive to the degree of depletion onto grain surfaces which, 
in turn, depends on the effective grain cross section.
 Another important parameter is the effective [C]/[O] ratio in the gas
 phase because this is critical for the formation of C-rich species
 such as C$_{3}$H$_{2}$, HC$_{3}$N, and \cch\, (see Terzieva \& Herbst 1998, 
Akyilmaz et al.~2007).  A [C]/[O] ratio close to unity can occur when
 a large fraction of the available oxygen condenses out in the form
 of water ice on grain surfaces. In this situation, the model 
 calculations suggest that CO takes the lion's share of both carbon
 and oxygen in the gas phase and that, as a result, little oxygen is
 left in atomic or molecular form.

  The results of Terzieva \& Herbst~(1998) show in particular that the \cch\, 
  abundance can change by an order of magnitude due either to the
  assumptions concerning [C]/[O] or due to assumptions concerning the
  rates of certain neutral-neutral reactions.  We find in contrast
  that the observed \cch\, abundance does not seem to  change
  greatly as a function of position and in fact, to a reasonable
approximation, \cch\, follows the \co\ distribution.  Where CO appears to be
depleted in L1498, \cch\, seems to be depleted also  and  this also seems
to be true for CB246 although the depletion holes are smaller. To some
extent, we presume that this reflects the fact that CO is the main source
of gas phase carbon in cores. When CO depletes, there is less carbon
available to form  species with 2 or more C atoms. It is striking also
that we derive very similar \cch\, abundances in the two cores
which we have studied. This may be due to chance 
 but it is interesting that our estimated \cch\, abundance is
similar to that derived in TMC1 and L183 (see Table 2 of Terzieva and
Herbst 1998). 

 It is also  worth noting
 that \cch, being a radical, is subject to destruction by atomic
oxygen and nitrogen.  Thus from the UMIST database (Woodall et al.~2007),
  one finds  that \cch\, reacts with O forming CO and CH at a rate of
  $1.7\times10^{-11}$ cm$^{3}$ s$^{-1}$ and the rate for atomic N is similar. 
  This compares with a rate for depletion onto dust grain surfaces of
  roughly $10^{-17}$n$_{\rm H}$ s$^{-1}$ where  n$_{\rm H}$ is the hydrogen nucleon
  density in cm$^{-3}$ and a rate of $10^{-9}$ cm$^{3}$ s$^{-1}$ for reactions
  with ions such as C$^{+}$ and H$_{3}^{+}$. For an ionization degree
  of $10^{-8}$, which is fairly typical in cores (see Fig. 5 in
  Walmsley et al.~2004), reactions with ions
  and depletion onto grain surfaces are competitive with one another. 
  Destruction by atomic oxygen and nitrogen  will dominate if the abundances
  of these species exceeds about $6\times10^{-7}$ relative to hydrogen. 
  The atomic abundances are  hence critical for the \cch\, abundance and it
  is presumably for this reason (at least as far as O is
  concerned) that  the results of model calculations show \cch\, to be 
  very sensitive to the gas
  phase [C]/[O] ratio (e.g. Terzieva and
  Herbst 1998, their Table 5). In fact, the only one of their model  
  predictions
  for \cch\, which
  compares reasonably with our observed values has a [C]/[O] ratio of 0.8
  (run 3 in their Table 5).  We conclude that high [C]/[O] gas phase ratios
  (and consequently a low atomic O abundance) are part of the explanation
  of the relatively high observed \cch\, abundance. This presumably implies
  a scenario with a large fraction of oxygen locked-up in the form of
  water ice on grain surfaces (see e.g. Hollenbach et al.~2009).

\subsection{C$_{2}$H as a magnetic field probe}
 As mentioned in the Introduction, \cch\ is potentially capable
of being used to measure the Zeeman effect. The CN(1$-$0) transition
has been successfully used for this purpose (Falgarone et al.~2008)
and superficially at least, \cch\  has
similar characteristics to CN (they are iso-electronic). 
We have therefore used the results of Crutcher et al.~(1996) together
with the Zeeman splitting calculations of Bel \& Leroy~(1998) to
infer the expected RMS sensitivity to magnetic field $\sigma_B$ of a \cch\ measurement 
at the intensity peak in L1498 for an integration time $t_{int}$. Thus we take

\begin{equation}
\sigma _{\rm B} \approx \frac{2}{Z}\, \frac{T_{\rm sys}}{T^{*}_{\rm A}}\frac{\Delta\nu_{\rm FWHM}}{\delta\nu\,t_{\rm int}}
\end{equation}
In this equation, we use our observed line intensities $T^{*}_{\rm A}$,
system temperature $T_{\rm sys}$, the spectral resolution $\delta\nu$
and width $\Delta \nu_{\rm FWHM}$ for L1498 as well as
the expected
Zeeman splitting factors $Z$, in Hz $\mu$G$^{-1}$, from  Bel and Leroy (1998) 
to derive expected values of $\sigma_{\rm B}$ for all six \cch(1$-$0) transitions. 
The results, for an hour of integration, are given in Table \ref{tab:zeeman} 
where we see component 5 of \cch\, is the most favorable for detecting 
Zeeman splitting and that, as for CN, there is a large variation of expected 
splitting between components. 
This latter property is of fundamental
importance for the purpose of eliminating instrumental effects
such as beam squint.

\begin{table}[!h]
\caption{Expected values of $\sigma_{\rm B}$ for all six \cch(1$-$0) 
transitions in L1498 at the offset $(60,-40)$ for an hour of integration and 
a spectral resolution equal to 40~kHz.}
\begin{center}
\begin{tabular}{ccccc}
\hline
comp. & $Z$ & $T^{*}_{\rm A}$ & $\Delta\nu_{\rm FWHM}$ & $\sigma_{\rm B}$\\
no. & [Hz $\mu$G$^{-1}$] & [K] & [kHz] & [$\mu$G]\\
\hline
1 & 2.6 & 0.9 & 80 & 740\\
2 & 0.7 & 1.4 & 100& 2210\\
3 & 2.3 & 1.0 & 90 & 860\\
4 & 0.93& 1.1 & 95 & 1990\\
5 & 2.8 & 1.1 & 85 & 590\\
6 & 0.93& 0.9 & 75 & 1860\\
\hline
\multicolumn{5}{l}{\tiny{See Table \ref{tab:c2h10p} for component labels.}}
\end{tabular}
\end{center}
\label{tab:zeeman}
\end{table}%
One sees also that without a considerable improvement in system temperature, 
it will be difficult to get below $3\sigma_{\rm B}$ limits of 200~$\mu$G. 
There have been claims of fields of this order however (e.g. Shinnaga et al.~1999); Crutcher~(1999)
 found a reasonable fit to the Zeeman data available at the time
 of $B_{\rm los}=80\times n_{4}^{0.46}$ $\mu$G
 with $n_{4}=10^{-4}\, n($H$_{2})$ which would suggest that fields of
 order 200 $\mu$G are reasonable in sources like L1498. We conclude
 therefore that fields of this order may be detectable using \cch\ but it
 is extremely difficult with current sensitivities.   On the other
 hand, in more general, prospects do not seem worse than with CN.


\section{Conclusions}

We have carried out a study of the behaviour of the abundance of \cch\ toward two starless cores with contrasting properties, L1498 and CB246. The main conclusions of our work are the following:

\begin{itemize}
\item[1.] In L1498, \cch\ shows a distribution similar to that observed 
in other species attributed to depletion onto grain surfaces in the 
central region; by contrast, in CB246 the dust and the \cch\ emission 
have similar distributions. 
\item[2.] The two cores show a clear signature for deviations from LTE 
populations in \cch: 
spectra in most positions deviate from expectations assuming LTE and thus
a single-temperature LTE model cannot fit the data.
\item[3.] There are positions of high \cch\ optical depth in both 
sources; in addition, L1498 shows 
self-absorption toward the dust peak. 
\item[4.] Our Monte 
Carlo model shows that the observed deviations from LTE 
can be qualitatively
understood, but reliable collisional rate calculations for
\cch\ are needed in order to make further progress.
\item[5.] The non-LTE deviations have not prevented from 
computing column density values based on LTE. We also found that the \cch
\ abundance relative to H$_{2}$ is remarkably constant outside regions of 
high CO depletion with a value of $(1.0\pm0.3)\times10^{-8}$ in L1498  
and $(0.9\pm0.3)\times10^{-8}$ in CB246. 
One possible implication is that the abundances of 
atomic oxygen and nitrogen are extremely low (below $6\times10^{-7}$ 
relative to H).
\item[6.] We derived a new set of frequencies for all the six hyperfine 
components of \cch(1$-$0) and seven components of \cch(2$-$1), computing 
an improved set of spectroscopic constants for \cch. 
\end{itemize}

\begin{acknowledgements}
This effort/activity is supported by the European Community Framework Programme 7, Advanced Radio Astronomy in Europe, grant agreement no.: 227290. M.P. and D.G. acknowledge support from the EC Research Training Network
MRTN-CT-2006-035890 ``Constellation: The Origin of Stellar Masses''.
H.S.P.M. is grateful for support by the Bundesministerium f\"ur Bildung und Forschung
(BMBF) administered through Deutsches Zentrum f\"ur Luft- und Raumfahrt (DLR). 
His support is aimed in particular at maintaining the CDMS.
\end{acknowledgements}



\begin{thebibliography}{9}



\bibitem[Aikawa et al.(2003)]{ao03}
Aikawa, Y., Ohashi, N. \& Herbst, E. 
2003, \apj, 593, 906

\bibitem[Aikawa et al.(2005)]{ah05}
Aikawa, Y., Herbst, E., Roberts, H. \& Caselli, P. 
2005, \apj, 620, 330

\bibitem[Akyilmaz et al.(2007)]{af07}
Akyilmaz, M., Flower, D. R., Hily-Blant, P., Pineau Des For\^ets, G. 
\& Walmsley, C. M. 
2007, A\&A, 462, 221

\bibitem[Andr\'e et al.(1996)]{aw96}
Andr\'e, P., Ward-Thompson, D. \& Motte, F. 
1996, A\&A, 314, 625

\bibitem[Bel \& Leroy (1998)]{bl98}
Bel, N. \& Leroy, B. 
1998 A\&A, 335, 1025

\bibitem[Bergin \& Tafalla(2007)]{bt07}
Bergin, E. A. \& Tafalla, M. 
2007, Ann. Rev. A\&A, 45, 339

\bibitem[Caselli et al.(1995)]{cm95}
Caselli, P., Myers, P. C. \& Thaddeus, P. 
1995, \apj, 455, L77

\bibitem[Caselli et al.(2002)]{cm02}
Caselli, P., Benson, P. J., Myers, P. C. \& Tafalla, M. 
2002, \apj, 572, 238

\bibitem[Chandrasekhar \& M\"unch(1950)]{cm50}
Chandrasekhar, S. \& M\"unch, G. 
1950, \apj, 111, 142

\bibitem[Clemens(1988)]{cb88}
Clemens, D. P. \& Barvainis, R. 
1988, \apjs, 68, 257

\bibitem[Codella \& Scappini(1998)]{cs98}
Codella, C. \& Scappini, F. 
1998, MNRAS, 298, 1092

\bibitem[Crutcher et al.(1996)]{ct96}
Crutcher, R. M., Troland, T. H., Lazareff, B. \& Kazes, I. 
1996, \apj, 456, 217

\bibitem[Crutcher(1999)]{c99}
Crutcher, R.~M. 
1999, \apj, 520, 706

\bibitem[Dame et al.(1987)]{du87}
Dame, T. M., Ungerechts, H., Cohen, R. S., de Geus, E. J., Grenier, I. A., 
May, J., Murphy, D. C., Nyman, L.-A. \& Thaddeus, P. 
1987, \apj, 322, 706

\bibitem[Ebenstein \& Muenter(1984)]{HCN_dipole_HFS} 
Ebenstein, W.~L., \& Muenter, J.~S. 
1984, \jcp, 80, 3989 

\bibitem[Elias(1978)]{e78}
Elias, J.~H. 
1978, \apj, 224, 857

\bibitem[Flower et al.(2006)]{fp06}
Flower, D. R., Pineau Des For\^ets, G. \& Walmsley, C. M. 
2006, A\&A, 456, 215

\bibitem[Falgarone et al.(2008)]{ft08}
Falgarone, E., Troland, T. H., Crutcher, R. M. \& Paubert, G. 
2008, A\&A, 487, 247

\bibitem[Goodman et al.(1993)]{gb93}
Goodman, A. A., Benson, P. J., Fuller, G. A. \& Myers, P. C. 
1993, \apj, 406, 528

\bibitem[Gottlieb et al.(1983)]{gg83}
Gottlieb, C. A., Gottlieb, E. W. \& Thaddeus, P. 
1983, \apj, 264, 740

\bibitem[Green \& Thaddeus(1974)]{gre74} 
Green, S. \& Thaddeus, P. 
1974, \apj, 191, 653 

\bibitem[Guilloteau \& Baudry(1981)]{gui81} 
Guilloteau, S. \& Baudry, A. 
1981, \aap, 97, 213 

\bibitem[Hily-Blant et al.(2008)]{hw08}
Hily-Blant, P., Walmsley, C. M., Pineau des For\^ets, G. \& Flower, D. 
2008, A\&A, 480, L5

\bibitem[Hollenbach et al.(2009)]{hk09}
Hollenbach, D., Kaufman, M. J., Bergin, E. A. \& Melnick, G. J. 
2009, \apj, 690, 1497

\bibitem[Kirk et al.(2006)]{kc06}
Kirk, J. M., Ward-Thompson, D. \& Crutcher, R. M. 
2006, MNRAS, 369, 1445

\bibitem[(1997)]{lh97}
Launhardt, R. \& Henning, T. 1997, A\&A, 326, 329

\bibitem[Lee et al.(2004)]{lb04}
Lee, J.-E., Bergin, E. A. \& Evans, N.J. II 
2004, \apj, 617, 360

\bibitem[Lemme et al.(1996)]{lw96}
Lemme, C., Wilson, T. L., Tieftrunk, A. R. \& Henkel, C. 
1996, A\&A, 312, 585

\bibitem[Lique et al.(2009)]{liq09} 
Lique, F., van der Tak, F.~F.~S., K{\l}os, J., Bulthuis, J. \& Alexander, M.~H. 
2009, \aap, 493, 557 

\bibitem[Monteiro \& Stutzki(1986)]{ms86}
Monteiro, T. S. \& Stutzki, J. 
1986, MNRAS, 221, P33

\bibitem[M{\"u}ller et al.(2000)]{mk00}
M\"{u}ller, H.~S.~P., Klaus, T. \& Winnewisser, G. 
2000, A\&A, 357, L65

\bibitem[M{\"u}ller et al.(2001)]{CDMS_1}
M{\"u}ller, H.~S.~P., Thorwirth, S., Roth, D.~A.,
\& Winnewisser, G.
2001, A\&A, 370, L49-L52

\bibitem[M{\"u}ller et al.(2005)]{CDMS_2}
M{\"u}ller, H.~S.~P., Schl{\"o}der, F., Stutzki, J.,
\& Winnewisser, G.
2005, J. Mol. Struct, 742, 215

\bibitem[Myers et al.(1996)]{mm96}
Myers, P. C., Mardones, D., Tafalla, M., Williams, J. P. \& Wilner, D. J. 
1996, \apjl, 465, L133

\bibitem[Myers \& Benson(1983)]{mb83}
Myers, P. C. \& Benson, P. J. 
1983, \apj, 266, 309

\bibitem[Ohashi(1999)]{o99}
Ohashi, N. 1999, in Proceedings of Star Formation 
1999, ed. T. Nakamoto (Nobeyama Radio Observatory), 129

\bibitem[Pagani et al.(2007)]{pb07}
Pagani, L., Bacmann, A., Cabrit, S. \& Vastel, C.
2007, A\&A, 467, 179

\bibitem[Pagani et al.(2009)]{pd09}
Pagani, L., Daniel, F. \& Dubernet, M.-L.
2009, A\&A, 494, 719

\bibitem[Preibisch et al.(1993)]{po93}
Preibisch, Th., Ossenkopf, V., Yorke, H.W. \& Henning, Th. 
1993, A\&A, 279, 577

\bibitem[Sastry et al.(1981)]{sh81} 
Sastry, K.~V.~L.~N., Helminger, P., Charo, A., Herbst, E., \& De Lucia, F.~C. 
1981, \apjl, 251, L119 

\bibitem[Schmid-Burgk et al.(2004)]{H13CO+_HFS} 
Schmid-Burgk, J., Muders, D., M{\"u}ller, H.~S.~P., \& Brupbacher-Gatehouse, B. 
2004, \aap, 419, 949 

\bibitem[Shinnaga et al.(1999)]{st99}
Shinnaga, H., Tsuboi, M. \& Kasuga, T. 1999, in Proceedings of Star Formation 
1999, ed. T. Nakamoto (Nobeyama Radio Observatory), 175

\bibitem[Shirley et al.(2005)]{sn95}
Shirley, Y. L., Nordhaus, M. K., Grcevich, J. M., Evans, N. J., II, 
Rawlings, J. M. C. \& Tatematsu, K. 
2005, \apj, 632, 982

\bibitem[Tafalla et al.(2002)]{tm02}
Tafalla, M., Myers, P. C., Caselli, P., Walmsley, C. M. \& Comito, C. 
2002, \apj, 569, 815

\bibitem[Tafalla et al.(2004)]{tm04}
Tafalla, M., Myers, P. C., Caselli, P. \& Walmsley, C. M. 
2004, A\&A, 416, 191

\bibitem[Tafalla et al.(2006)]{tm06}
Tafalla, M., Santiago-Garc\'ia, J., Myers, P. C., Caselli, P., Walmsley, C. M. 
\& Crapsi, A. 
2006, A\&A, 455, 577
 
\bibitem[Tassoul(1978)]{t78}
Tassoul, J.-L. 
1978, Theory of Rotating Stars (Princeton: Princeton Univ. Press)
 
\bibitem[Terzieva \& Herbst(1998)]{th98}
Terzieva, R. \& Herbst, E. 
1998, \apj, 501, 207
 
\bibitem[Thum et al.(2008)]{tw08}
Thum, C., Wiesemeyer, H., Paubert, G., Navarro, S. \& Morris, D. 
2008, PASP, 120, 777
 
\bibitem[Turner et al. (1999)]{tur99} 
Turner, B.~E., Terzieva, R. \& Herbst, E. 
1999, \apj, 518, 699 

\bibitem[Walmsley et al. (2004)]{wf04}
Walmsley, C.~M., Flower, D.~R. \& Pineau des For\^ets, G.
2004, A\&A, 418, 1035

\bibitem[Woodall et al.(2007)]{wa07}
Woodall, J., Ag\'undez, M., Markwick-Kemper, A. J. \& Millar, T. J. 
2007, A\&A, 46, 1197



\end{thebibliography}
\end{document}